\documentclass[lineno]{jfm}
\usepackage{verbatim}
\usepackage{graphicx}
\usepackage{newtxtext}
\usepackage{newtxmath}
\usepackage{natbib}
\usepackage{hyperref}
\hypersetup{
    colorlinks = true,
    urlcolor   = blue,
    citecolor  = black,
}

\newcommand{\RomanNumeralCaps}[1]
\linenumbers
\usepackage{siunitx}
\usepackage[justification=justified]{caption}

\title{Generalizable super-resolution turbulence reconstruction from minimal training data}

\author{
  Haokai Wu$^{1}$\thanks{These authors contributed equally to this work.},
  Yong Cao$^{1,2}$\footnotemark[1]
  \corresp{\email{yongcao@sjtu.edu.cn}},
  Yaoran Chen$^{3}$,
  Shujin Laima$^{4}$,
  Wenli Chen$^{4}$,
  Dai Zhou$^{1,2}$,
  \and Hui Li$^{4}$
}

\affiliation{
\aff{1}School of Ocean and Civil Engineering, Shanghai Jiao Tong University, Shanghai, PR China
\aff{2}State Key Laboratory of Ocean Engineering, Shanghai Jiao Tong University, Shanghai, PR China
\aff{3}School of Future Technology, Shanghai University, Shanghai, 200444, PR China
\aff{4}School of Civil Engineering, Harbin Institute of Technology, Harbin, PR China}

\begin{document}
\maketitle
\begin{abstract}
Fully resolving turbulent flows remains challenging due to turbulent systems' multiscale complexity. Existing data-driven approaches typically demand expensive retraining for each flow scenario and struggle to generalize beyond their training conditions. Leveraging the universality of small-scale turbulent motions (Kolmogorov’s K41 theory), we propose a Scale-oriented Zonal Generative Adversarial Network (SoZoGAN) framework for high-fidelity, zero-shot turbulence generation across diverse domains. Unlike conventional methods, SoZoGAN is trained exclusively on a single dataset of moderate-Reynolds-number homogeneous isotropic turbulence (HIT). The framework employs a zonal decomposition strategy, partitioning turbulent snapshots into subdomains based on scale-sensitive physical quantities. Within each subdomain, turbulence is synthesized using scale-indexed models pre-trained solely on the HIT database. SoZoGAN demonstrates high accuracy, cross-domain generalizability, and robustness in zero-shot super-resolution of unsteady flows, as validated on untrained HIT, turbulent boundary layer, and channel flow. Its strong generalization, demonstrated for homogenous and inhomogenous turbulence cases, suggests potential applicability to a wider range of industrial and natural turbulent flows. The scale-oriented zonal framework is architecture-agnostic, readily extending beyond GANs to other deep learning models.

\end{abstract}



\section{Introduction}
\label{sec:headings}

Turbulence plays a fundamental role across numerous industrial and natural systems, from aerospace engineering to atmospheric dynamics and oceanographic flows. While homogeneous turbulence features spatially uniform statistics, non-homogeneous turbulence - such as wall-bounded flows - exhibits statistical variations driven by boundaries or mean flow gradients. Despite these differences, a central tenet of turbulence theory, Kolmogorov’s K41 hypothesis \citep{K41} and its subsequent extensions \citep{Benzi1993}, asserts the universality of small-scale turbulent motions at sufficiently high Reynolds numbers. Specifically, statistical properties in the inertial and dissipation ranges become universal: independent of large-scale flow structures and controlled solely by the mean energy dissipation rate $\epsilon$ and the kinematic viscosity $\nu$. Later studies have further demonstrated that this universality extends beyond velocity statistics to velocity gradient fields. For even-order moments (particularly second-order) of velocity gradients, robust universality is observed across homogenous and non-homogeneous turbulence \citep{Buaria2025}. Notably, extreme events associated with velocity gradients maintain universal characteristics among various flow configurations, providing a bridge between homogenous and non-homogeneous turbulent regimes \citep{Buaria2025,Buaria2021,Schumacher2014}.

Despite profound theoretical insights, fully resolving turbulent flows remains a formidable challenge. Experimental methods face fundamental spatiotemporal resolution limits \citep{bib5}, while numerical simulations demand rapidly escalating computational resources with increasing Reynolds number. Direct Numerical Simulation (DNS) of homogeneous isotropic turbulence requires grid counts scaling as $N \sim \mathrm{Re}_{\lambda}^{9/4}$ \citep{bib1}, with even steeper scaling in wall-bounded flows. Large-Eddy Simulation (LES) of wall-bounded turbulence imposes significant burdens with $N \sim \mathrm{Re}^{13/7}$ \citep{Choi2012}. Such scaling renders high-fidelity simulations infeasible for many practical, high-Reynolds-number flows \citep{bib6,bib7}.

Recent advances in deep learning have introduced promising data-driven approaches to these challenges by inferring high-resolution turbulence features from sparsely sampled or low-resolution data. These methods learn nonlinear mappings between coarse flow variables and multiscale structures using large datasets, achieving notable success in reconstructing detailed turbulent fields \citep{bib9,bib10,bib11,bib30,manohar2018data,lozano2023machine,vinuesa2022enhancing}. Among these, super-resolution frameworks—originally developed for image processing—aim to produce high-resolution flow fields $\mathbf{q}_{\rm HR} \in \mathbb{R}^{n_x \times n_y}$ from coarse inputs $\mathbf{q}_{\rm LR} \in \mathbb{R}^{(n_x/r) \times (n_y/r)}$, where $r$ is the downsampling factor. These models infer unresolved turbulent motions from coarse data, bridging the gap between limited-resolution simulations or experiments and the full turbulence spectrum.

Deterministic neural networks, such as Convolutional Neural Networks (CNNs) \citep{bib12,bib31,bib32} and Graph Neural Networks (GNNs) \citep{Han2022PredictingPI}, have been pioneering tools in turbulence super-resolution due to their ability to extract local and non-Euclidean features, respectively. However, their deterministic nature limits their capacity to capture the inherently stochastic, high-dimensional dynamics of turbulence, especially in regimes dominated by chaotic small-scale fluctuations \citep{Du2024NC}.

More recently, generative models have brought a paradigm shift by integrating probabilistic learning into turbulence super-resolution. Super-resolution Generative Adversarial Networks (SRGANs) \citep{bib11, bib13, bib14, yasuda2023spatio} - including Wasserstein GANs (WGANs) \citep{GAO2022111270}, conditional GANs (cGANs) \citep{Mirza2014dfp} and CycleGAN \citep{Kim2021} - as well as diffusion-based models \citep{Dhariwal2021, Gao10204802}, reconstruct microscale turbulence in a statistically consistent manner through adversarial or hierarchical training. Extensions of this line of research include the integration of super-resolution into subgrid-scale modelling for very coarse LES data, thereby linking deep learning-based reconstruction with turbulence modelling theory through cGAN (supervised) and CycleGAN (unsupervised) frameworks \citep{Maejima2025}. Notably, SRGANs preserve turbulent energy spectra and dissipation mechanisms, enhancing physical fidelity \citep{bib13, bib16, Kim2024}. Hybrid frameworks combining normalizing flows and GNNs further enable probabilistic generation of instantaneous turbulent fields \citep{sun2023unifying}. Although such methods reconcile deterministic approximations with turbulent stochasticity, their training remains computationally expensive, particularly when resolving multiscale interactions.

Two major limitations persist in current data-driven turbulence super-resolution methods. First, existing models exhibit limited generalization ability beyond their training datasets; adapting to new turbulence conditions often requires costly retraining or fine-tuning with high-fidelity data \citep{bib23,bib24,bib29}. Second, achieving broad generalizability typically demands extensive, high-resolution, multi-condition datasets \citep{bib28}, whose generation imposes substantial computational burdens and hinders scalability. These challenges arise primarily due to the vast variability of flow patterns and boundary effects encountered in practical applications, which are often not well represented in training data. Recent efforts have tackled the training-data scarcity problem by leveraging efficient sampling strategies. Notably, \citet{bib23} proposed a multi-scale extraction approach, in which subdomains of various vortical sizes are cropped from a single instantaneous high-resolution flow field to construct a diverse set of training samples across scales. This method exploits the spatial scale similarity inherent in turbulence to synthetically enrich the training dataset, thereby substantially alleviating the need for multiple training datasets. 

To address two longstanding challenges in turbulence super-resolution—limited generalization and high data dependence—this work leverages the universality concept of small-scale turbulent motions to propose the scale-oriented zonal GAN (SoZoGAN). This innovative framework enables efficient, high-fidelity super-resolution by generalizing across diverse flow regimes using only coarse input fields.
SoZoGAN operates in a ``zero-shot’’ manner: it predicts small-scale turbulent structures in unseen flow configurations without retraining or labeled data from the new target tasks \citep{bib15}. Unlike conventional approaches, SoZoGAN is pretrained exclusively on a single, readily accessible dataset: homogeneous isotropic turbulence (HIT) at moderate Reynolds number. Leveraging a carefully designed scaling transformation of HIT data, we construct a library of scale-specific SRGAN models that span a wide range of turbulent scales, conceptually akin to the multi-scale extraction approach of \citet{bib23}. To adapt to spatially inhomogeneous flows, the framework partitions target fields into subdomains based on scale-sensitive physical quantities. Each subdomain is then super-resolved using the pretrained SRGAN model corresponding to its characteristic scale. By synergistically coupling offline scale-specific pretraining with online zero-shot synthesis, SoZoGAN delivers robust cross-domain performance. Furthermore, the scale-oriented zonal strategy is model-agnostic and can be integrated flexibly with diverse deep learning architectures beyond GANs.

\begin{figure}
\centering
\includegraphics[width=1.03\textwidth]{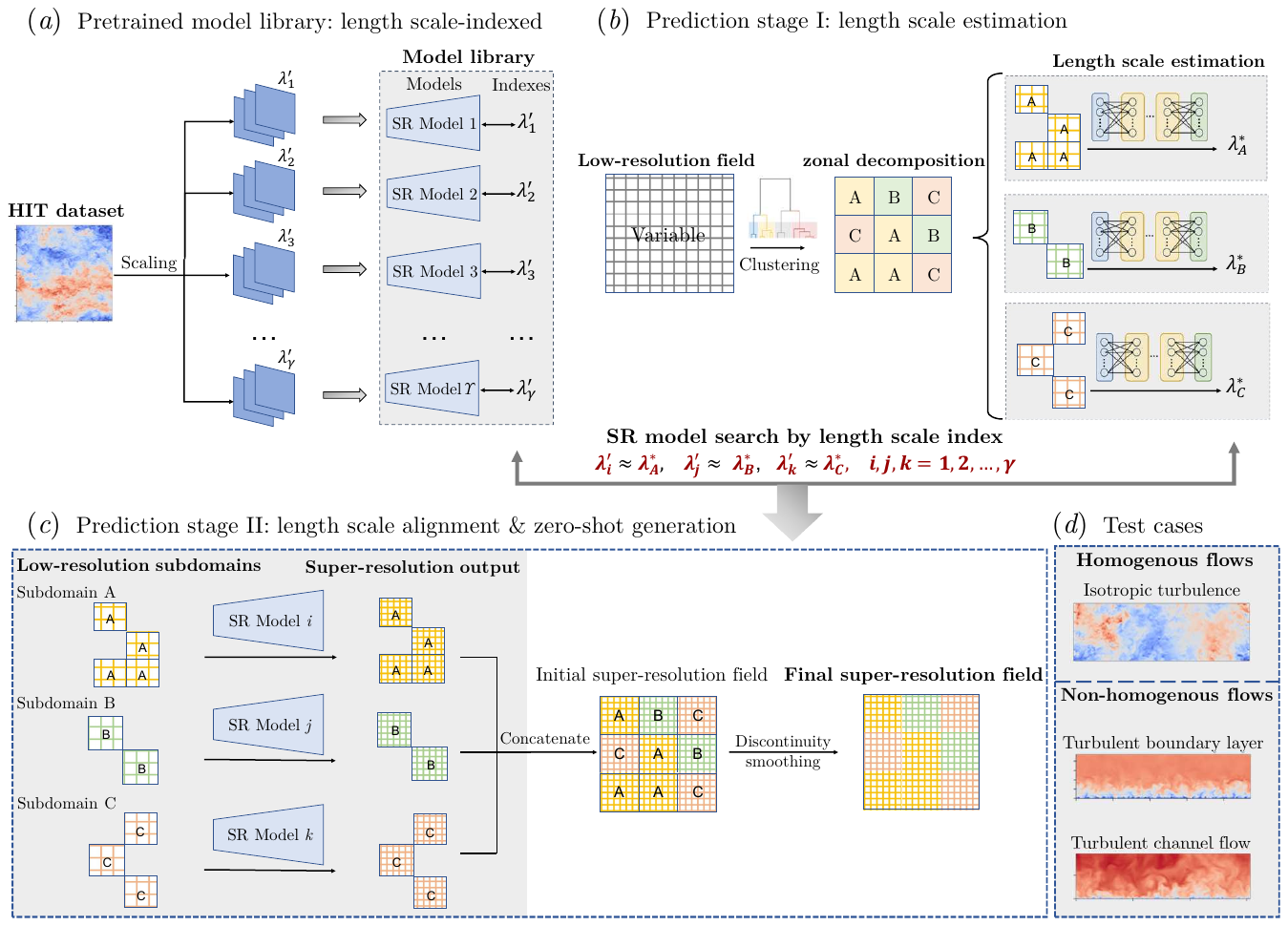}
\caption{The SoZoGAN framework for turbulence super-resolution that generalizes across diverse turbulent flows using ``zero-shot'' transfer. (a) Establishment of a pretrained super-resolution model library using a single dataset. (b) Zonal decomposition and microscale estimation based on the low-resolution flow field. (c) Microscale alignments of the decomposed subdomains and the procedure of ``zero-shot'' generation from low-resolution subdomains to the super-resolution global flow field. (d) Test cases of the proposed framework, including HIT, turbulent boundary layer and channel flow. }\label{fig1}
\end{figure}

\section{Methodology}\label{Methodology}

\subsection{Overview}

Turbulent flows in practical scenarios often exhibit pronounced anisotropy and inhomogeneity due to disturbances originating either within the flow or near boundaries. These effects cause the characteristic turbulent scales to vary significantly across different spatial regions of the global flow field. Such spatial scale variability poses a major challenge for turbulence super-resolution (SR), because a single SR model—trained under fixed-scale assumptions—cannot simultaneously recover flow structures across all local regions with high physical fidelity. Consequently, an adaptive SR framework is needed to (i) identify locally varying physical scales, and (ii) apply scale-appropriate generative models without retraining for each new case.

To address this challenge, we propose the Scale-oriented Zonal Generative Adversarial Network (SoZoGAN), a ``zero-shot'' SR framework designed for high-fidelity reconstruction of inhomogeneous turbulence without additional training or fine-tuning for new flow scenarios. The central idea is to exploit the universality of small-scale turbulent motions, learned from a single homogeneous isotropic turbulence (HIT) dataset, and transfer this SR capability to diverse target flows through a scale-oriented decomposition and model-alignment workflow. As illustrated in Figure~\ref{fig1}, SoZoGAN operates in three connected stages, each with a clear motivation:

\begin{enumerate}
    \item \textbf{Microscale-indexed model pretraining} (figure~\ref{fig1}(a)):  
    Starting from HIT data, a carefully designed scaling transformation generates training samples covering a wide range of microscale conditions. For each distinct microscale, a dedicated SRGAN is pretrained, forming a library of scale-specific models indexed by their microscale values. This offline stage is designed to encapsulate scale‑dependent generative characteristics into specialized models, thereby eliminating the need for retraining in subsequent target cases (see Section~\ref{subsec1}).
    
    \item \textbf{Zonal decomposition and microscale estimation} (figure~\ref{fig1}(b)):  
    The low-resolution (LR) target field is partitioned into subdomains with approximately uniform scale features, based on scale-sensitive quantities such as velocity fluctuations, velocity gradients, or spectral content. An MLP-based estimator then predicts the characteristic microscale of each subdomain directly from LR inputs. This stage helps to align each subdomain to the most suitable pretrained SRGAN in a physically grounded way, preventing global heterogeneity from biasing local scale identification (see Section~\ref{subsec2}).

    \item \textbf{Scale-oriented zero-shot generation} (figure~\ref{fig1}(c)):
    For each subdomain, the SRGAN whose microscale index best matches the estimated local scale is retrieved from the library. The subdomains are super-resolved individually without retraining, and the resulting SR patches are blended to reconstruct a globally continuous field. This stage aims to deploy specialized priors where they fit best, while ensuring seamless global integration through blending and continuity handling (see Section~\ref{subsec3}).    
\end{enumerate}

By coupling offline scale-indexed pretraining with online scale-aware deployment, SoZoGAN bridges the gap between single-dataset training and generalized SR across spatially heterogeneous turbulence, enabling zero-shot reconstructions for previously unseen flow configurations.

\subsection{Pretrained model library}\label{subsec1}

Homogeneous and isotropic turbulence (HIT) is the most fundamental and widely applicable type of turbulence. In terms of its generation mechanism, the multi-scale similarity observed in isotropic turbulence is also present in other turbulence types, all of which adhere to the energy cascade theory. This provides a theoretical foundation for data-driven models trained on isotropic turbulence data to be ``zero-shot'' transferred to different turbulence types. Therefore, in the pretraining stage, the public database of DNS HIT \citep{bib19} made available by the Johns Hopkins Turbulence Databases (JHTDB) is employed to exclusively pretrain the SR models.

To generate turbulent flow fields with spatially varying scale characteristics, we require a set of SR models, each pretrained to learn a corresponding structural pattern at a specific characteristic scale (microscale). 
The pretraining data, which capture distinct turbulence features at these characteristic scales, are manually extended from the single HIT dataset using a physics-related scaling transformation. Each SR model is pretrained on specific data that represent a microscale and is indexed accordingly with the corresponding scale value, as illustrated in figure~\ref{fig1}(a).

In the final step, these pretrained SR models are organized into a model library. This library facilitates efficient model retrieval during the following ``zero-shot'' generation stage, enabling the appropriate model to be selected based on the required scale characteristics. The implementation details of data curation and model pretraining are provided below.

\subsubsection{Data curation}

The pretraining HIT database was simulated within a $x\times y\times z = 2\pi\times 2\pi\times 2\pi$ box using pseudo-spectral method, with a grid size of $8192\times 8192\times 8192$. Five three-dimensional snapshots are collected in this database with $\mathrm Re_{\lambda}\sim1200$. In each snapshot, 3 velocity components of $u$, $v$ and $w$ are included for $x$, $y$ and $z$ directions.

\begin{figure}
\centering
\includegraphics[width=1.0\textwidth]{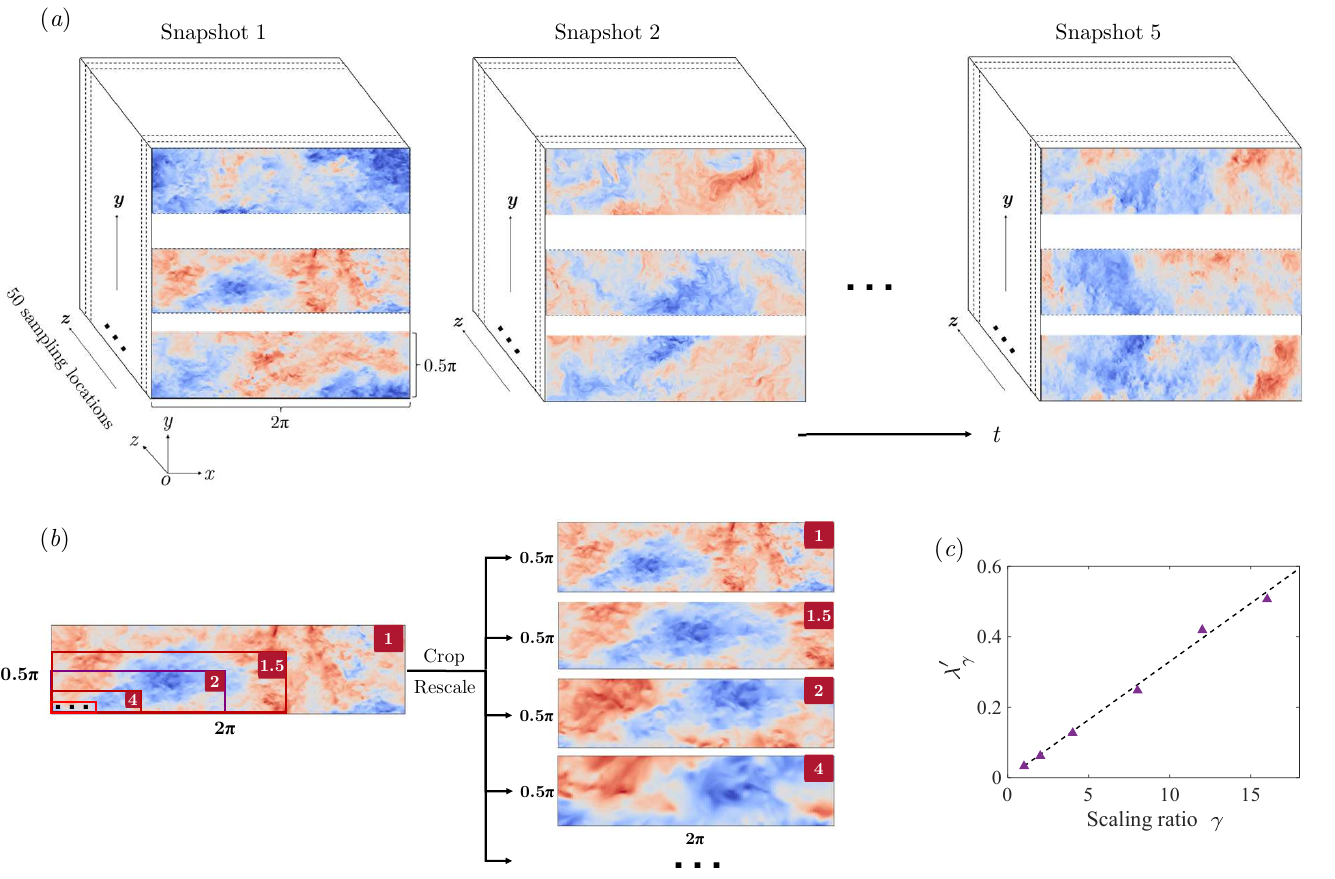}
\caption{Data curation from a single HIT dataset and verification of the scaling transformation method: (a) Schematic of the spatial and temporal sampling procedure to obtain initial planes, followed by (b) extension into velocity sub-planes. Scaling transformation is applied to extract sub-plane samples with different microscales from the initial planes. The numbers labeled in the top-right corner of each sub-plane ($\gamma$) represent the scaling ratios in the $x$ or $y$ direction relative to the width or height of the initial plane. (c) Validity of the scaling transformation. The scaling ratios $\gamma$ and the Taylor microscales $\lambda'_{\gamma}$ are nearly in a linear relationship for a series of HIT sub-planes.}\label{fig2jfm}
\end{figure}

Since this study focuses on two-dimensional spatial super-resolution, pretraining samples of velocity distribution planes are extracted from each snapshot, as illustrated in figure~\ref{fig2jfm}(a). To obtain these samples, we first extract 50 equally spaced $x$–$y$ slices along the $z$-direction for each snapshot, ensuring uniform coverage of streamwise–spanwise structures throughout the HIT data. Within each extracted $x$–$y$ slice, we partition along the $y$-direction into three non-overlapping rectangular planes, each with a size of $x \times y = 2\pi \times 0.5\pi$ (grid resolution $8192 \times 2048$). These partitioned planes constitute our initial sample planes. Applying this procedure to all five snapshots yields a total of $50 \times 3 \times 5 = 750$ initial planes for pretraining.

In our pretraining stage, for scale-oriented learning of a set of SR models, HIT samples with varying Taylor microscales are artificially extracted from a single original dataset. Leveraging the self-similarity of turbulence, a scaling method is applied to crop the initial sampled plane ($2\pi\times0.5\pi$) proportionally in both the $x$ and $y$ directions, creating multiple sub-planes in figure~\ref{fig2jfm}(b) (grid resolution $8192/\gamma\times 2048/\gamma$). The $\gamma$ ($\gamma\ge1$) labeled in the top-right corner of each sub-plane represents the scaling ratio in the $x$ or $y$ direction relative to the width or height of the initial sampled plane. Simultaneously, the space length of each sub-plane should be rescaled to the initial plane ($2\pi\times0.5\pi$) to reproduce the larger-scale distribution characteristics from the cropped smaller-scale ones. This operation purposely filters out smaller-scale turbulent structures from the original HIT data, thereby preserving larger-scale turbulence features in the extended sub-plane samples. Through a series of scaling transformations, pretraining samples with diverse feature scales are obtained.

To assess the validity of this scaling transformation, it is necessary to examine the relationship between the scaled physical properties (specifically, the Taylor microscale) and the scaling ratio $\gamma$. The Taylor microscales of sub-plane samples with varying $\gamma$ are calculated as ${\lambda_\gamma'}^2 = 2 \left\langle u^2 \right\rangle / \left\langle \left[ \partial u / \partial y \right]^2 \right\rangle$~\citep{bib1}, where the coordinate $y$ is consistent with that of the initial sampled plane. The operator $\left\langle \cdot \right\rangle$ denotes the statistical mean over 750 sub-plane samples at each corresponding $\gamma$. As shown in figure~\ref{fig2jfm}(b), $\gamma$ and $\lambda_\gamma'$ are generally positively correlated in a linear relationship. It indicates that ${\lambda_\gamma'} \approx \gamma \cdot \lambda_1'$. Therefore, by adjusting $\gamma$ based on $\lambda_1'$ ($\sim 0.0331$) of the original HIT field, we can easily obtain the corresponding pretraining samples for the desired Taylor microscale of ${\lambda_\gamma'}$.

\subsubsection{Pretrained SRGAN models}

Unlike conventional supervised regression models that rely solely on specific network architectures, such as multilayer perceptrons (MLPs) \citep{bib33} or convolutional neural networks (CNNs) \citep{bib35}, a generative adversarial network (GAN) is a training framework that can be instantiated with different architectures. In image super-resolution tasks, both generator and discriminator components of a GAN are often implemented as CNNs, adversarially trained to enhance the fidelity of fine-scale structures. Such adversarial training is particularly effective at recovering high-frequency content and complex microscale details in the reconstructed fields.

In the SoZoGAN framework, all SR models follow the super‑resolution GAN (SRGAN) design proposed by \citet{bib17}, where the generator ($G$) and discriminator ($D$) are jointly trained within an adversarial learning paradigm. Both $G$ and $D$ employ CNN‑based architectures. The detailed architectures of $G$ and $D$ follow prior deep-learning-based SR studies \citep{bib17,bib18,bib16}.

Let the high-resolution (HR) reference flow field be $\mathbf{q}_{\mathrm{HR}} \in \mathbb{R}^{n_x \times n_y}$, and the corresponding LR input be $\mathbf{q}_{\mathrm{LR}} \in \mathbb{R}^{(n_x/r) \times (n_y/r)}$, where $r$ is the downsampling factor. In the supervised pretraining stage, $\mathbf{q}_{\mathrm{LR}}$ is obtained from $\mathbf{q}_{\mathrm{HR}}$ via a downsampling (average pooling) operator $\mathcal{D}$, i.e., $\mathbf{q}_{\mathrm{LR}} = \mathcal{D}(\mathbf{q}_{\mathrm{HR}})$. The adversarial training of SRGAN is formulated as:
\begin{equation}
\min_{G} \max_{D} V(D, G) =
\mathbb{E}_{\mathbf{q}_{\rm HR} \sim p_{\mathrm{data}}}\,[\log D(\mathbf{q}_{\rm HR})] +
\mathbb{E}_{\mathbf{q}_{\rm LR} \sim p_{n}}\,[\log (1 - D(G(\mathbf{q}_{\rm LR})))] ,
\end{equation}
where $p_{\mathrm{data}}$ and $p_{n}$ denote the probability distributions of HR and LR samples, respectively. It is important to note that HR data $\mathbf{q}_{\mathrm{HR}}$ are required only during the pretraining stage, where they serve as supervised targets for learning both $G$ and $D$. Once the scale-indexed SRGAN library has been trained, the zero-shot prediction stage operates solely on LR inputs $\mathbf{q}_{\mathrm{LR}}$ without any need for HR references. In inference, $G$ produces SR reconstructions $\hat{\mathbf{q}}_{\mathrm{HR}} = G(\mathbf{q}_{\mathrm{LR}})$, recovering small-scale turbulent structures using the scale-specific representations acquired during pretraining.

During pretraining, the generator $G$ in each SRGAN is optimized with paired LR and HR fields. The training objective integrates pixel-wise accuracy, adversarial error, and physical consistency via a composite loss:
\begin{equation}
L_G = L_{\mathrm{MSE}} + \alpha\, L_{\mathrm{Adver}} + \beta\, L_{\mathrm{Conti}},
\end{equation}
where:
  \begin{equation}
    L_{\mathrm{MSE}} = \| \mathbf{q}_{\mathrm{HR}} - G(\mathbf{q}_{\mathrm{LR}}) \|_2^2,
  \end{equation}
  penalizing the $\ell_2$ error between the SR reconstruction and its HR reference,
  
  \begin{equation}
    L_{\mathrm{Adver}} = - \log\big(D(G(\mathbf{q}_{\mathrm{LR}}))\big),
  \end{equation}
  where $D(\cdot)$ estimates the probability that a sample belongs to the HR distribution \citep{bib13}, encouraging $G$ to produce perceptually realistic outputs, and
  
  \begin{equation}
    L_{\mathrm{Conti}} = \left\| \frac{\partial u}{\partial x} + \frac{\partial v}{\partial y} + \frac{\partial w}{\partial z} \right\|_{2,\mathrm{SR}},
  \end{equation}
  where $L_{\mathrm{Conti}}$ denotes the $\ell_2$ norm of the divergence of the reconstructed SR velocity field $(u, v, w)$, computed using finite‑difference approximations. This term enforces the incompressible continuity constraint during training. To calculate $\partial w/\partial z$ in $L_{\mathrm{Conti}}$, the SRGAN should also generate the SR $w$‑component on the neighboring $x$–$y$ plane, following the strategy employed by \citet{bib16}.

The balancing factors $\alpha$ and $\beta$ control the influence of the adversarial and physics-based terms; $\alpha$ is fixed to $10^{-3}$ following \citet{bib13}, while $\beta$ is adapted so that $L_{\mathrm{Conti}}$ constitutes more than 50\% of $L_G$ on average, ensuring a strong emphasis on physical consistency. As noted above, HR data $\mathbf{q}_{\mathrm{HR}}$ are used solely for pretraining; in the subsequent zero-shot prediction stage, $G$ operates exclusively on LR inputs $\mathbf{q}_{\mathrm{LR}}$ without requiring HR references.

To construct HR–LR training pairs, we treat all pretraining samples derived from the HIT dataset as HR velocity fields. Each HR field has dimensions $8192/\gamma \times 2048/\gamma \times 3$ (corresponding to the three velocity components) and is associated with a specific Taylor microscale $\lambda_\gamma'$. The corresponding LR inputs are generated by applying average pooling with downsampling factor $r$, where $r$ also corresponds to the desired super-resolution ratio. This produces LR fields of dimensions $8192/(\gamma\cdot r) \times 2048/(\gamma\cdot r) \times 3$. We therefore build a library of pretrained $r\times$ SRGAN models, with each model uniquely indexed by its Taylor microscale $\lambda_\gamma'$. 

Average pooling here is both computationally efficient and physically interpretable as an explicit low‑pass filtering process. In our study, the generated LR data are obtained through a well‑controlled pooling that preserves large‑scale, energy‑containing structures and part of the inertial‑range dynamics, while filtering out unresolved motions. This makes it suitable for obtaining LR data not only from numerical simulations but also from experimental measurements such as particle image velocimetry or multi‑fidelity regridded datasets \citep{Scarano2013, bib30, bib10}. Compared with our previous work \citep{bib16}, which applied an additional low‑pass filter after pooling to mimic small‑scale energy dissipation in very coarse‑grid simulations, the present study omits this step to solely considerate the resolution‑scale effects. Nonetheless, the EC‑SRGAN model proposed by \citet{bib16} can be integrated into the current framework to form a ``SoZoEC‑SRGAN'' variant for targeted small‑scale reconstruction in very coarse‑grid simulation scenarios. Extended work will specifically address these extreme scenarios in Section~\ref{subsec37}.

\subsection{Physics-guided zonal decomposition and microscale estimation}\label{subsec2}

The generation of small scales in inhomogeneous turbulence poses a fundamental challenge: how can we recover local small-scale structures when the input field exhibits dramatic spatial variations in its characteristic scales? Traditional super-resolution approaches, which treat the entire domain uniformly, inevitably fail when confronted with such heterogeneity. 
To address this limitation, we propose decomposing the low-resolution input field into physically-informed subdomains, where characteristic microscales are estimated or pre-calibrated within each subdomain to guide subsequent turbulence generation.

\subsubsection{Zonal decomposition through hierarchical clustering}

We first illustrate the zonal decomposition process using wall-bounded turbulence, a canonical example of strongly inhomogeneous flow. Such flows are typically homogeneous in the horizontal directions but exhibit pronounced variations in turbulence statistics along the wall-normal direction. This anisotropy calls for a decomposition strategy that explicitly respects spatial differences in turbulent scales.

In the SoZoGAN, zonal decomposition is not an independent preprocessing step but a prerequisite for physics-based microscale estimation. Partitioning the domain into subdomains with relatively uniform scale properties ensures that the estimated microscale parameters are representative of each subdomain, rather than being distorted by mixed-scale statistics. Without this step, parameter estimation would be less accurate, undermining the effectiveness of scale-specific SR model selection. To achieve this, we employ a hierarchical clustering approach based on scale-sensitive physical quantities extracted from the LR input field (figure~\ref{fig1}(b), middle panel). This method produces subdomains within which turbulence scales remain consistent, enabling physically meaningful subdomains for targeted SR reconstruction. The procedure for wall-bounded turbulence proceeds in four systematic steps:

First, we divide the coarse flow field into local grid groups, each containing points at the same wall-normal height, where turbulence statistics are approximately comparable. These groups form the fundamental units for hierarchical clustering.

Second, for each group, we compute scale-sensitive physical quantities from the coarse input field. Specifically, the one-dimensional wavenumber spectra of all three velocity components are calculated along the statistically homogeneous horizontal direction. These spectra describe the distribution of turbulent kinetic energy across spatial scales at each wall-normal location and serve as feature vectors for clustering.

While one-dimensional wavenumber spectra are used in this study for their clear physical interpretation and ability to represent the full turbulence scale distribution, the proposed zonal decomposition is not limited to this choice. Other scale-sensitive quantities—such as velocity gradient magnitude, turbulent kinetic energy (TKE), or eddy size from autocorrelation lengths—can also serve as clustering features, and multiple quantities may be combined if appropriately normalized. For the wall-bounded turbulence considered here, the one-dimensional spectra are chosen because they (i) link directly to turbulence scales, revealing both dominant and dissipative ranges; (ii) exploit directional homogeneity to reduce noise and isolate genuine wall-normal scale variations; and (iii) offer richer multi-scale detail than single-point measures, enabling more discriminative subdomain clustering.
 
Third, similarity between groups is measured as the Euclidean distance between their spectral feature vectors, giving $m(m-1)/2$ pairwise distances for $m$ groups. Hierarchical clustering with an average linkage criterion is then applied, iteratively merging wall-normal locations with the highest spectral similarity. The average distance between two merged clusters $C_1$ and $C_2$ is defined as:
\begin{equation}
d(C_{1}, C_{2})=\frac{1}{|C_{1}||C_{2}|} \sum_{x_{1} \in C_{1}} \sum_{x_{2} \in C_{2}} d(x_{1}, x_{2}),
\end{equation}
where $|C_{1}|$ and $|C_{2}|$ are the numbers of groups in each cluster, and $d(x_{1}, x_{2})$ is the Euclidean distance of the spectral feature vectors between local grid groups $x_{1}$ and $x_{2}$. This criterion promotes the formation of clusters with coherent scale characteristics.

While Euclidean distance is adopted here for its computational simplicity, numerical stability, and compatibility with the average-linkage criterion, the zonal decomposition can readily incorporate other similarity measures. For example, if spectral features are normalized as probability distributions, distribution-based metrics such as the Kullback–Leibler divergence or the Wasserstein-2 distance \citep{lienen2024} can be used to capture differences in the shape of the energy spectrum.

Finally, the hierarchical clustering process produces a dendrogram (``tree-shaped structure'') in which each local grid group starts as an individual leaf node and, at each iteration, the pair with the smallest average inter-cluster distance is merged into a new parent node \citep{Patel2015,Zhao2005}. This process continues until a single root is reached, with the vertical axis of the dendrogram representing the distance at which merges occur. The final decomposed subdomains are obtained by “cutting” this tree horizontally at a specified maximum inter-cluster distance: a smaller threshold preserves finer wall-normal scale variations by producing more subdomains, whereas a larger threshold yields fewer, coarser partitions. By tuning this distance, one can balance computational efficiency against the need to capture spatial variations in turbulence scales. The resulting adaptive decomposition maintains horizontal homogeneity within each layer while retaining the wall-normal variation of scales, ensuring that each subdomain is well suited for targeted super-resolution.

Although the above steps are presented for wall-bounded turbulence, the hierarchical clustering is generally applicable to other non-homogeneous turbulence types by adapting the definition of the fundamental units. For example, in atmospheric turbulence, where significant inhomogeneity exists in both longitudinal and latitudinal directions, the fundamental unit can be defined as a rectangular region in the plane (figure~\ref{fig1}(b)). The size of this rectangle determines the granularity of the zonal decomposition - smaller rectangles can better capture local turbulence variability. This flexibility makes the approach broadly adaptable to diverse flow configurations.

\subsubsection{Physical basis for microscale estimation}

The Taylor microscale $\lambda$ has a unique position in turbulence theory, marking the crossover between the energy-containing and dissipation ranges of the turbulent cascade~\citep{bib1}. It characterizes the length scale over which small-scale velocity fluctuations remain correlated, making it a central anchor for super-resolution reconstruction. However, coarse-grained fields typically capture only the large-scale motions, leaving microscale information severely underrepresented or entirely absent.

The key insight is to exploit the intrinsic relationships that govern turbulent scales. According to Kolmogorov's similarity theory, the Taylor microscale is fundamentally linked to large-scale energetics through the energy cascade mechanism. For isotropic turbulence, this relationship takes the form~\citep{bib1}:
\begin{equation}
\lambda / L=\sqrt{10} \mathrm{Re}_{L}^{-1/2},\quad \mathrm{Re}_{L}\equiv k_t^{1/2} L/\nu 
\label{eq:1}
\end{equation}
where $\mathrm{Re}_{L}$ is the large-scale Reynolds number, $L$ is the integral length scale characterizing the energy-containing eddies, $k_t=\frac{3}{2}\overline{u'^2}$ is the turbulent kinetic energy, $u'$ is the root-mean-square velocity fluctuation, and $\nu$ is the kinematic viscosity.

Equation~\eqref{eq:1} reveals a profound physical truth: even when microscale structures are not visible in coarse data, their characteristic scales remain encoded in the macroscale flow properties. This relationship suggests that a mapping function $\mathcal{F}: \lambda = f(L, u', \nu)$ may exist for more general turbulent flows. To approximate this mapping for complex inhomogeneous turbulence, we employ a multilayer perceptron (MLP) architecture, as illustrated in figure~\ref{fig1}(b). This neural network serves as a physical anchor, translating available macroscale information from coarse inputs into estimates of the embedded microscales. The following Section~\ref{MLP} provides details on its input–output mapping, training datasets, loss formulation, and optimization strategy.

\subsubsection{Microscale estimation via pretrained MLP}\label{MLP}

Once the flow field is decomposed into physically coherent subdomains, each subdomain requires microscale calibration to guide the selection of appropriate super-resolution models. Our MLP architecture, featuring three hidden layers with eight neurons each, undertakes this mapping from macroscale flow parameters to embedded Taylor microscales ($\mathcal{F}: \lambda = f(L, u', \nu)$), as depicted in figure~\ref{fig1}(b).

The MLP is trained using a diverse collection of homogeneous isotropic turbulence databases \citep{bib20,bib21,bib22,bib19}, as detailed in table~\ref{tab1}. These datasets span a broad range of flow conditions, enabling the network to learn varied scale relationships. The macroscale quantities—integral length scale $L$, RMS velocity fluctuation $u'$, and kinematic viscosity $\nu$—are extracted from low-resolution fields and serve as the network input. These three quantities together determine the large-scale Reynolds number $Re_{L}$, which, under isotropic turbulence scaling, relates to the Taylor Reynolds number as $Re_{\lambda}\propto Re_{L}^{1/2}$. Through this link, variations in $(L, u', \nu)$ implicitly encode variations in $Re_{\lambda}$, giving the MLP a physically grounded basis to potentially extrapolate to higher-$Re$ cases. The corresponding Taylor microscales $\lambda_{\rm DNS}$, obtained from high-resolution DNS data, serve as the regression output targets. We adopt the mean squared error (MSE) between the predicted microscale $\lambda_{\mathrm{pred}}$ and the DNS-derived reference $\lambda_{\mathrm{DNS}}$ as the loss function for the MLP:
\begin{equation}
    L_{\mathrm{MLP}} = \frac{1}{N_{\mathrm{HIT}}} \sum_{i=1}^{N_{\mathrm{HIT}}} \left[ \lambda_{\mathrm{pred}}^{(i)} - \lambda_{\mathrm{DNS}}^{(i)} \right]^2,
\end{equation}
where $N_{\mathrm{HIT}}$ denotes the number of training samples, and $i$ is the sample index.

\begin{table}
  \begin{center}
\def~{\hphantom{0}}
  \begin{tabular}{lccccc}
	Data sources & $\mathrm{Re}_{\lambda}$ & $\lambda$ & $L$ & $u'$ & $\nu$\\
	\cite{bib20} & 186   & 0.0890  & 0.5600  & 0.2300  & $1.100	\times 10^{-4}$  \\
	\cite{bib21} & 418   & 0.1127  & 1.364  & 0.6860  & $1.850	\times 10^{-4}$ \\
	\cite{bib22} & 610   & 0.0674  & 1.392  & 1.569  & $1.732	\times 10^{-4}$ \\
        \cite{bib19}   & 1200   & 0.0331  & 1.186  & 1.573  & $4.385	\times 10^{-5}$ \\
  \end{tabular}
  \caption{Flow parameters of different HIT databases used for MLP pre-training.}
  \label{tab1}
  \end{center}
\end{table}

To enhance the robustness of the MLP and prevent overfitting to any single turbulence configuration, we augment the training dataset using the scaling transformations described in Section~\ref{subsec1}. This artificial augmentation broadens the parameter coverage, allowing the network to generalize beyond the specific conditions found in the training datasets. The ADAM optimizer~\citep{bib36} is used to train the MLP, with early stopping criteria applied to avoid overfitting and to ensure the learned mapping reflects genuine physical relationships instead of dataset-specific artifacts.

The pre-trained MLP serves as a bridge between the observable macroscale characteristics in coarse input fields and the latent microscale features essential for accurate super-resolution. During the prediction stage, when the MLP is applied to more complex turbulent flows, such as wall-bounded turbulence, the required input $L$ is evaluated locally within each fundamental unit employed by the hierarchical-clustering-based zonal decomposition (figure~\ref{fig1}(b)). In each such unit, $L$ is obtained from the velocity-field data by computing the longitudinal two-point autocorrelation function $R_{uu}(r)$ along a chosen homogeneous direction and integrating it according to the standard definition:
\begin{equation}
    L = \int_0^{r_0} R_{uu}(r)\,dr,
\end{equation}
where $r_0$ denotes the first zero-crossing point of $R_{uu}$. For wall-bounded turbulence, the fundamental units correspond to layers at distinct wall-normal heights, and input $L$ is computed separately within each layer. For other turbulence types with inhomogeneity in different directions, each fundamental unit is taken as a small rectangular region in the plane, and input $L$ is calculated independently in each rectangle (figure~\ref{fig1}(b), right). By providing ``zero-shot'' microscale estimations locally, the MLP enables the selection of the most appropriate super-resolution model from the pre-trained SR library, ensuring that turbulence generation remains faithful to the local flow physics.

\subsection{Scale-oriented ``zero-shot'' prediction}\label{subsec3}

Following the zonal microscale estimation, the Taylor microscales $\lambda$ predicted by the MLP at diverse fundamental units are spatially averaged within each decomposed subdomain. This averaging yields the averaged microscale $\lambda^*$ for each subdomain, as illustrated in figure~\ref{fig1}(b). The value of $\lambda^*$ is then used to characterize the typical microscale flow structures that should be generated within the corresponding subdomain.

To ensure scale fidelity, we select optimal super-resolution models for each subdomain from our pretrained library through microscale alignment. These SR models that are trained on HIT samples with known Taylor microscales are chosen according to a alignment criterion: the microscale $\lambda^*$ for each subdomain is matched as closely as possible to the characteristic microscales $\lambda_{\gamma}'$ of the available SR models. In particular, for subdomains A, B, and C shown in figure~\ref{fig1}(b), the estimated microscales satisfy $\lambda_A^* \approx \lambda'_{i}$, $\lambda_B^* \approx \lambda'_{j}$, and $\lambda_C^* \approx \lambda'_{k}$, where the subscripts $(A,B,C)$ denote subdomain identifiers and $(i,j,k)$ index the corresponding pretrained models from our library. Based on the selected pretrained SRGAN models, small-scale turbulence in each subdomain is generated via zero-shot inference, without any additional training or fine-tuning (shown in the gray region in figure~\ref{fig1}(c)). The applicability of this zero-shot is constrained by the smallest resolvable turbulence scale in the input data and by Reynolds-number compatibility, as discussed in Section~\ref{limitations}.

After super-resolving each subdomain individually, the generated subdomains are reassembled according to their original spatial arrangement, forming a composite global super-resolution flow field. However, independent generation in separate subdomains may introduce discontinuities or inconsistencies at the subdomain interfaces. To address this, we employ a postprocessing step using a $1\times$ SRGAN model. This model operates at identical grid resolution between input and output and is also pretrained (without any additional retraining) exclusively on lightweight HIT sub-planes extracted from the original high-resolution database. The $1\times$ model serves to repair and smooth potential discontinuities in the global field, but does not generate new small-scale turbulence. As a result, the final super-resolved flow (shown in the right portion of figure~\ref{fig1}(c)) achieves both physical fidelity within each subdomain and global coherence across the entire domain, all accomplished without any retraining or optimization during test stage.

\section{Results and discussions}\label{Results}

To assess SoZoGAN's capacity for cross-domain generalization, we evaluate its performance exclusively through zero-shot transfer to unseen turbulent configurations: homogeneous isotropic turbulence at $\mathrm{Re}_{\lambda} \approx 433$ (distinct from training Reynolds numbers) and inhomogeneous wall-bounded flows including turbulent boundary layers (TBL) and channel flows (figure~\ref{fig1}(d)). These test cases, all sourced from the JHTDB for rigorous DNS benchmarking, probe the model's ability to reconstruct fine-scale turbulent structures across fundamentally different physical scenarios without any additional training. The evaluation encompasses both qualitative flow visualization and quantitative statistical analysis, with all low-resolution inputs generated through average pooling from the corresponding high-fidelity DNS fields.

\subsection{SoZoGAN's generalization}
To rigorously assess the cross-domain generalization ability of SoZoGAN, we conduct a comprehensive quantitative evaluation involving multiple types of turbulence. We consider several accuracy metrics: root mean square error (RMSE), coefficient of determination ($R^2$) averaged across all velocity components, and higher-order statistics—up to fourth-order moments—of both velocity fluctuations and gradients (table \ref{tab2}). Across this diverse set of flow cases, SoZoGAN demonstrates remarkable zero-shot transfer performance. The model achieves $R^2$ values close to one, minimal RMSE, and errors in the second to fourth moments of velocity consistently below 5\%. Even though the errors for velocity gradient moments are moderately higher (averaging around 14\% for second to fourth-order moments), they remain well within the acceptable benchmarks established by \citet{Buaria2023}. It is important to note that \citet{Buaria2023} examined only HIT at varying Reynolds numbers. In contrast, our study covers both a broader range of flow types and a wider spectrum of Reynolds numbers, creating a more stringent test of generalization.

\begin{table}
  \begin{center}
\def~{\hphantom{0}}
  \begin{tabular}{lccc}
      Flow types & HIT & TBL & Channel flow \\[3pt]
      RMSE & 0.0675   & 0.0082  & 0.0079   \\
      $R^2$ & 0.9777   & 0.9201  & 0.9876   \\
      Variance $\widetilde{u}$ error & 4.84\%   & 3.03\%  & 1.44\%   \\
      Skewness $\widetilde{u}$ error & 0.24\%   & 1.37\%  & 2.73\%   \\
      Flatness $\widetilde{u}$ error & 1.85\%   & 1.22\%  & 0.98\%   \\
      Variance $\partial \widetilde{u}/\partial y$ error & 15.08\%   & 8.88\%  & 7.41\%   \\
      Skewness $\partial \widetilde{u}/\partial y$ error & 9.04\%   & 14.03\%  & 23.19\%   \\
      Flatness $\partial \widetilde{u}/\partial y$ error & 11.02\%  & 20.58\%  & 13.20\%   \\
  \end{tabular}
  \caption{Performance metrics of SoZoGAN applied to turbulence super-resolution in different flow types.}
  \label{tab2}
  \end{center}
\end{table}

\subsection{Homogeneous turbulence at unseen Reynolds numbers}

We assess SoZoGAN's generalization capability for homogeneous isotropic turbulence through zero-shot transfer from the training Reynolds number ($\mathrm{Re}_\lambda \approx 1200$) to an unseen testing condition ($\mathrm{Re}_\lambda \approx 433$). The evaluation dataset comprises 1000 DNS snapshots from the JHTDB at $\mathrm{Re}_\lambda \approx 433$, where each snapshot provides all three velocity components ($u$, $v$, $w$) on a uniform grid. To maintain consistency with pretraining data, we extract $400 \times 400$ planar velocity fields from each snapshot, covering a physical domain of size $2\pi \times 0.5\pi$ in the $(x,y)$-plane.

Leveraging HIT's statistical homogeneity, we employ a simplified SoZoGAN variant that operates without zonal decomposition. Model selection is achieved by matching the test case's Taylor microscale to our library's precomputed values (figure~\ref{fig1}(a)). The $1/r \times$ low-resolution input is generated via an average pooling for high-resolution snapshots with a factor of $r$, preserving large-scale structures while reducing spatial detail.

\begin{figure}
\centering
\includegraphics[width=0.85\textwidth]{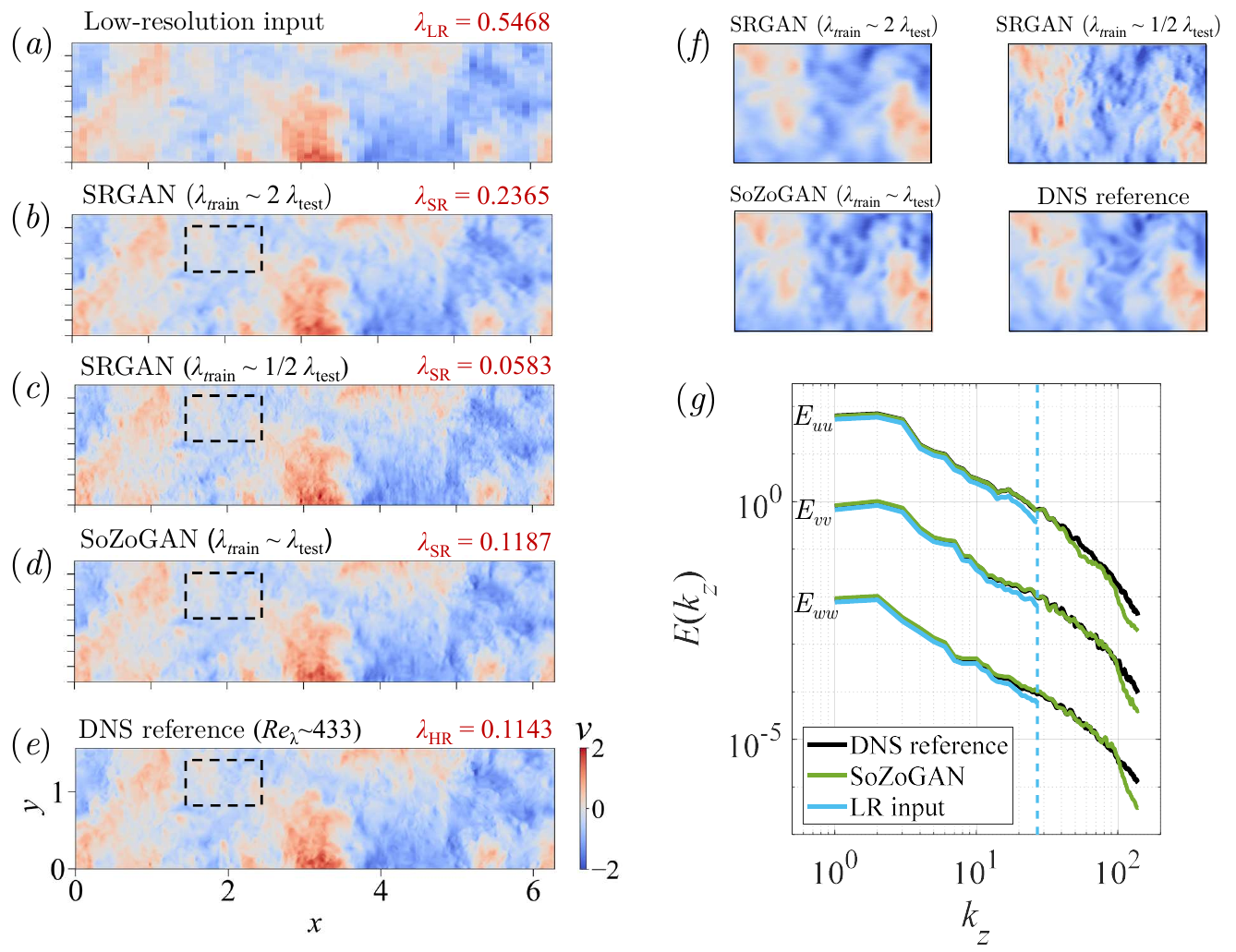}
\caption{Effect of training-testing scale alignment on small-scale HIT generations with SoZoGAN. (a) Low-resolution ($1/5 \times$) input field for the $v$-velocity component. (b)-(d) Instantaneous $v$-component fields generated by SRGANs trained at different Taylor microscales: (b) training scale much larger than testing, (c) testing scale much larger than training, (d) matched scales, compared with (e) high-resolution reference fields. Taylor microscales $\lambda_{\text{LR}}$, $\lambda_{\text{SR}}$, and $\lambda_{\text{HR}}$ associated with input, reconstructed, and DNS reference fields are marked along with the corresponding instantaneous fields. (f) Enlarged views of the boxed regions in (b)-(e) highlight differences in generated vortex structures relative to the high‑resolution reference. (g) Wavenumber spectra for all velocity components ($u, v, w$) comparing SoZoGAN generations (aligned case), coarse inputs, and DNS references. $E_{uu}$ is scaled by $1\times10^4$, $E_{vv}$ by $1\times10^2$, and $E_{ww}$ remains unscaled for clarity. The dash line represents the cut-off wavenumber of the LR input.}\label{fig3jfm}
\end{figure}

To demonstrate the critical role of scale alignment, we systematically evaluate model performance under both matched and mismatched training--testing conditions. Figure~\ref{fig3jfm}(b)-(d) compares instantaneous $v$-component fields generated by SRGANs trained at different microscales from identical $1/5\times$ low-resolution inputs (figure~\ref{fig3jfm}(a)), while figure~\ref{fig3jfm}(e) shows the corresponding DNS reference. The Taylor microscales, calculated from these velocity fields and indicated in each panel's upper-right corner, are denoted as $\lambda_{\text{LR}} = 0.5468$ (input), $\lambda_{\text{SR}}$ (predictions), and $\lambda_{\text{HR}} = 0.1143$ (DNS reference). The coarse-grained input exhibits an microscale approximately five times larger than the DNS reference (figure~\ref{fig3jfm}(e)), significantly attenuating small-scale information. When the SRGAN is trained at substantially larger microscales ($\lambda_{\text{train}} \approx 2\lambda_{\text{test}}$), it captures mainly large-scale structures while failing to reconstruct fine details, producing predictions with microscales twice the reference value (figure~\ref{fig3jfm}(b)). Conversely, models trained at smaller microscales ($\lambda_{\text{train}} \approx 0.5\lambda_{\text{test}}$) generate artificial fine-scale fluctuations, yielding unphysically smaller characteristic scales at the given $\mathrm{Re}_{\lambda}$. Optimal performance occurs when the microscales of the pretraining and testing data are well aligned (figure~\ref{fig3jfm}(d)). As highlighted in the enlarged views (figure~\ref{fig3jfm}(f)) of the boxed regions in figure~\ref{fig3jfm}(b)-(e), SoZoGAN accurately reconstructs small-scale turbulence from $1/5\times$ inputs, with vortical structures closely matching DNS results.

The effect of proper scale alignment is therefore striking: when the microscales in training and testing are matched, SRGAN faithfully reconstructs small-scale turbulent structures, with the predicted microscale remaining within 5\% of the DNS value. This result confirms that scale alignment is essential for producing physically consistent super-resolution. Figure~\ref{fig3jfm}(g) further supports this observation by presenting wavenumber spectra for all three velocity components $(u, v, w)$ based on SoZoGAN reconstructions. By comparing these spectra to those of both the low-resolution input and the high-fidelity DNS, it becomes clear that SoZoGAN is able to restore not only the inertial subrange but also dissipation-range spectral content far beyond what is present in the coarse input fields.

\subsection{Inhomogeneous turbulence: turbulent boundary layer}

Compared to homogeneous turbulence, inhomogeneous wall-bounded turbulence poses a more rigorous challenge for SoZoGAN due to its dramatic variation in characteristic scales along the wall-normal direction. In turbulent boundary layers, wall shear generates intense small-scale vortices near the wall, while large-scale structures dominate the outer region, where turbulence gradually decays toward laminar conditions. This strong scale heterogeneity highlights the need for adaptive models capable of accurately capturing multiscale features.

\subsubsection{Preprocessing of testing data}

The testing data here are sourced from a TBL database generated through DNS of incompressible flow over a no-slip flat plate. Detailed descriptions of the computational domain and grid resolutions can be found in the previous studies \cite{bib25}. The coordinate system is defined by $x$, $y$, and $z$ representing the streamwise, wall-normal, and cross-flow directions respectively, with the corresponding velocity components denoted as $u$, $v$, and $w$. 

The focus of our analysis is a particular two-dimensional cross-plane ($y–z$ plane) at a momentum thickness-based Reynolds number $\mathrm Re_{\theta}\sim1000$. From this cross-plane, we extract data for the three velocity components over a size of $y\times z = 23.65L \times 94.6L$ ($L$ is half thickness of the plate). The corresponding grid resolution is $400\times 400$, evenly distributed along each direction. To obtain low-resolution datasets, the high-resolution data are downsampled with the factors of $r=[5,10,16]$, reducing the grid size to $80\times 80$, $40\times 40$ and $25\times 25$, respectively. This reduction is achieved via average pooling for each velocity component \citep{bib34}. To ensure the statistical reliability, 1000 snapshots are collected at equal time intervals over an adequately long period, providing a testing dataset of low-resolution inputs.

\begin{figure}
\centering
\includegraphics[width=0.85\textwidth]{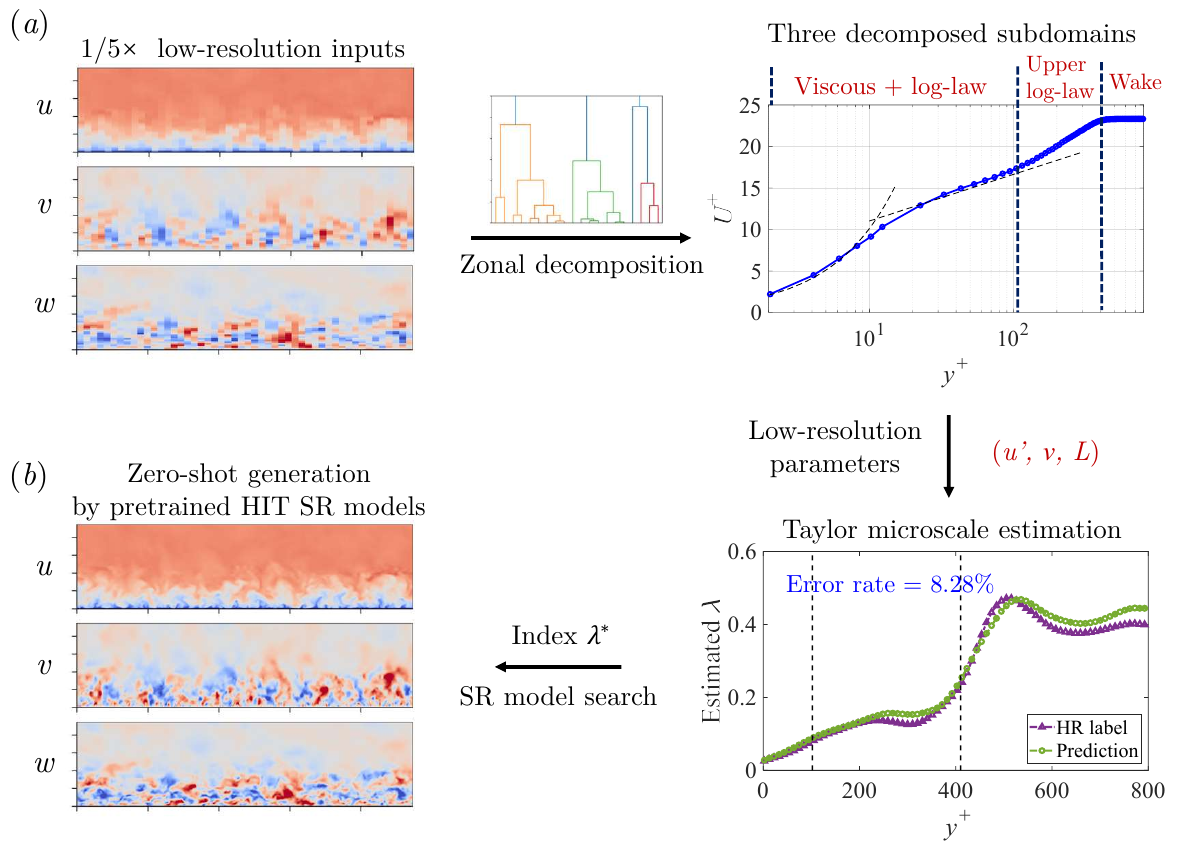}
\caption{Zonal decomposition and Taylor microscale estimation in TBL super-resolution using SoZoGAN. (a) Hierarchical clustering partitioning the low-resolution TBL domain into three physically meaningful subdomains, with interfaces at $y^+ \approx 102$ and $y^+ \approx 410$, corresponding to the viscous plus lower log-low, upper log-law, and wake regions, respectively. (b) Wall-normal profiles of the estimated Taylor microscale ($\lambda$) predicted by the MLP based on macroscale parameters from low-resolution input, compared with the high-resolution DNS reference.}\label{fig5jfm}
\end{figure}

In the prediction stage, spatial differences in scale distribution should be first identified from the low-resolution field. We perform the zonal decomposition using hierarchical clustering guided by scale-sensitive wavenumber spectra. For the TBL considered, statistical homogeneity in the horizontal directions allows subdomain boundaries to be placed only along the wall-normal direction, simplifying the decomposition process. The number of subdomains is user-defined: increasing it allows finer adaptation to local scale variations and can enhance reconstruction accuracy, but at the cost of requiring a larger library of pretrained SR models. This trade-off will be examined in Section~\ref{Robustness}. As a representative case, figure~\ref{fig5jfm}(a) presents a hierarchical clustering result that partitions the domain into three subdomains, with physical interfaces located at $y^+ \approx 102$ and $y^+ \approx 410$, where $y^+$ denotes the wall-normal distance normalized by the viscous length scale $\delta_{\nu}$. These interfaces approximately demarcate (1) viscous plus lower log-low, (2) upper log-law, and (3) wake region (large-scale motions dominate), reflecting classic boundary layer distribution and demonstrating the physical validity of our zonal decomposition strategy.

After zonal decomposition, the MLP, pretrained on the universal macroscales-to-microscales relationship learned from HIT data, is applied directly to the low-resolution TBL field to estimate the local Taylor microscale ($\lambda$), as illustrated in figure~\ref{fig5jfm}(b). Accurate estimation of the Taylor microscale enables scale precalibration, which is critical for scale-oriented super-resolution in inhomogeneous turbulence. At each wall-normal location ($y^+$), the MLP takes as input the macroscale parameters $L$, $\nu$, and $u'$, all of which are well-preserved and accessible even in coarsened fields. The MLP-estimated wall-normal profile of the Taylor microscale is compared against values obtained from high-resolution DNS, showing close agreement with a low spatially averaged error of $8.28\%$. When spatially averaged over each decomposed subdomain, the estimated $\lambda^*$ values are 0.0505, 0.1509, and 0.4192 from the near-wall to the outer subdomains, respectively. These averages are used to identify the best-matching microscale indices in the model library, guiding the selection of the appropriate SR models for each individual subdomain.

\subsubsection{Roles of zonal scale alignment}

\begin{figure}
\centering
\includegraphics[width=1.0\textwidth]{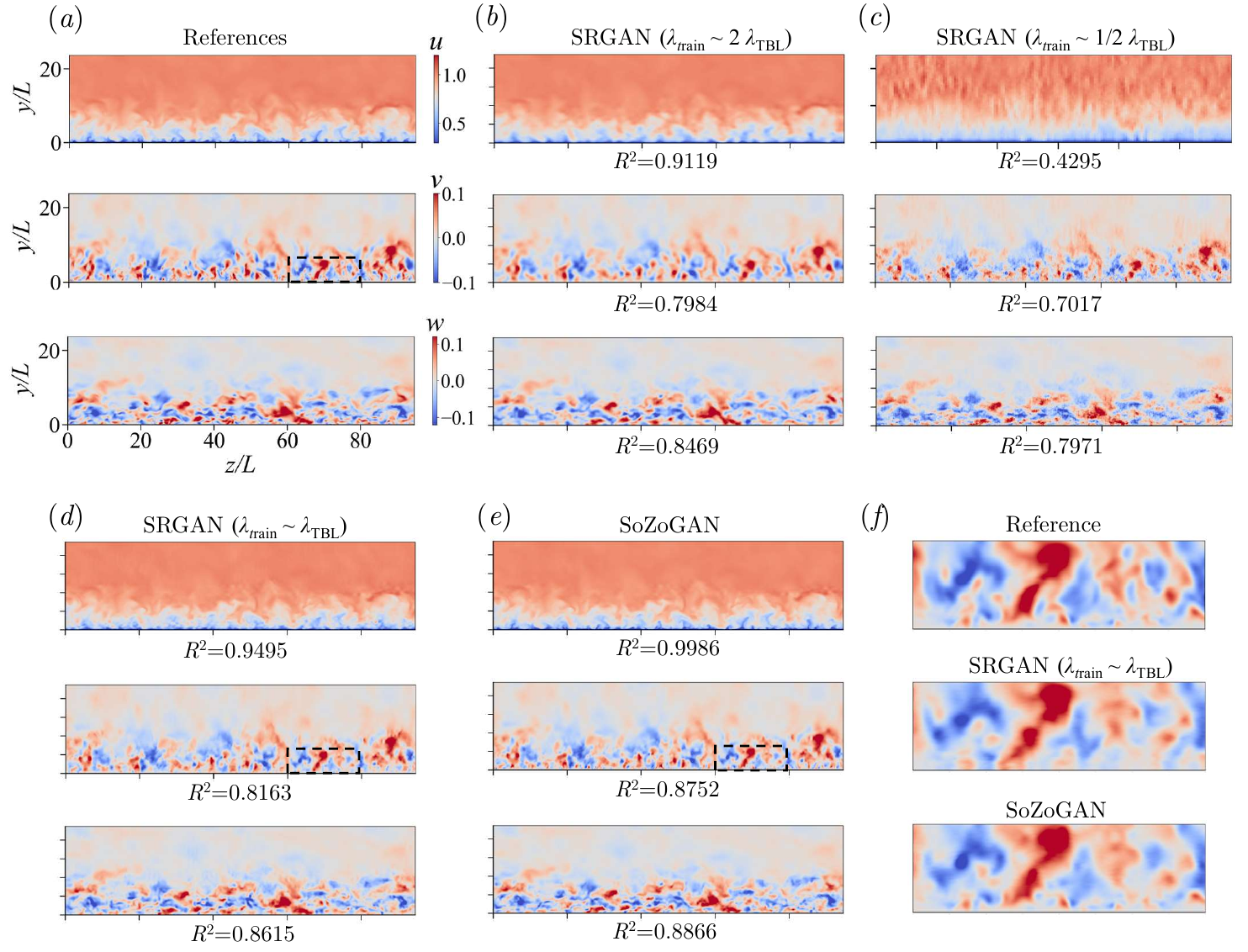}
\caption{Effects of microscale alignment and zonal generation and on TBL super-resolution. (a)-(e) Instantaneous fields of three velocity components (b)-(d) globally generated by SRGANs and (e) zonally generated by SoZoGAN, with $R^2$ values shown below each field. (f) Local zoom‑in views of the black dashed boxes shown in (a), (d) and (e), highlighting the generated near‑wall small‑scale structures of the SRGAN ($\lambda_{\rm train}\approx\ \lambda_{\rm TBL}$) and SoZoGAN, compared with the DNS reference field.}\label{fig6jfm}
\end{figure}

We take $5 \times$ super-resolution as our primary case to dissect the roles of microscale alignment and zonal generation within the proposed SoZoGAN framework. To this end, we conduct an ablation study, the results of which are shown in figure~\ref{fig6jfm}. Starting from $1/5 \times$ low-resolution inputs (Figure~\ref{fig5jfm}(a)), we compare the instantaneous super-resolved velocity fields for all three components generated by different SRGAN-based approaches against the corresponding DNS reference (Figure~\ref{fig6jfm}(a)) and SRGAN baseline (Figure~\ref{fig6jfm}(f)) snapshots.

Taylor microscale alignment is first evaluated using three SRGAN models pretrained at different scales (figure \ref{fig6jfm}(b)-(d)). Figure~\ref{fig6jfm}(b) illustrates the result obtained by applying a sole $5 \times$ SRGAN model, pretrained on HIT samples with a Taylor microscale ($\lambda_{\rm train}$) approximately twice that of the target TBL test data ($\lambda_{\rm TBL} \sim 0.2321$, spatially averaged). This model reproduces large-scale turbulence characteristics but fails to resolve the smaller-scale structures near the wall, as it overestimates the relevant microscale features. Conversely, figure~\ref{fig6jfm}(c) presents results from an SRGAN model pretrained with a much smaller scale index ($\lambda_{\rm train}\approx\ 1/2\lambda_{\rm TBL}$). Here, the super-resolved fields display pronounced non-physical fluctuations, especially in the $u$ component, indicating that the model has transferred small-scale features inconsistent with the actual microscales of the TBL, thereby generating unrealistic reconstructions. Figure~\ref{fig6jfm}(d) demonstrates the effect of applying microscale alignment: a sole $5 \times$ SRGAN model is chosen from the library such that its training microscale matches the spatially averaged $\lambda_{\rm TBL}$ of the target field ($\lambda_{\rm train}\approx\ \lambda_{\rm TBL}$). The resulting super-resolved fields exhibit substantially reduced non-physical fluctuations compared to the previous cases and closely match the distributions seen in the DNS reference, as quantified by improved $R^2$ values. This outcome confirms that microscale precalibration enables reliable transfer of turbulence characteristics from HIT data to TBL reconstructions.

Next, we examine how zonal generation enhances reconstruction quality. Figure~\ref{fig6jfm}(e) displays results from the full SoZoGAN framework, which combines scale alignment and zonal generation. Compared to the sole scale-aligned model, zonal generation markedly enhances the reproduction of fine velocity fluctuations, especially in the near-wall region (as highlighted in figure~\ref{fig6jfm}(f)). This improvement is reflected in the accuracy metrics, with $R^2$ reaching nearly 0.9 for $v$ and $w$ and approaching 1 for $u$. This zonal strategy effectively addresses wall turbulence's spatial heterogeneity by applying scale-matched models to specific flow subdomains. The study reveals that successful turbulence reconstruction depends on two key elements: first, exact microscale matching to ensure basic physical validity, and second, intelligent zonal decomposition to handle the spatially varying nature of wall turbulence characteristics.

\begin{figure}
\centering
\includegraphics[width=1.0\textwidth]{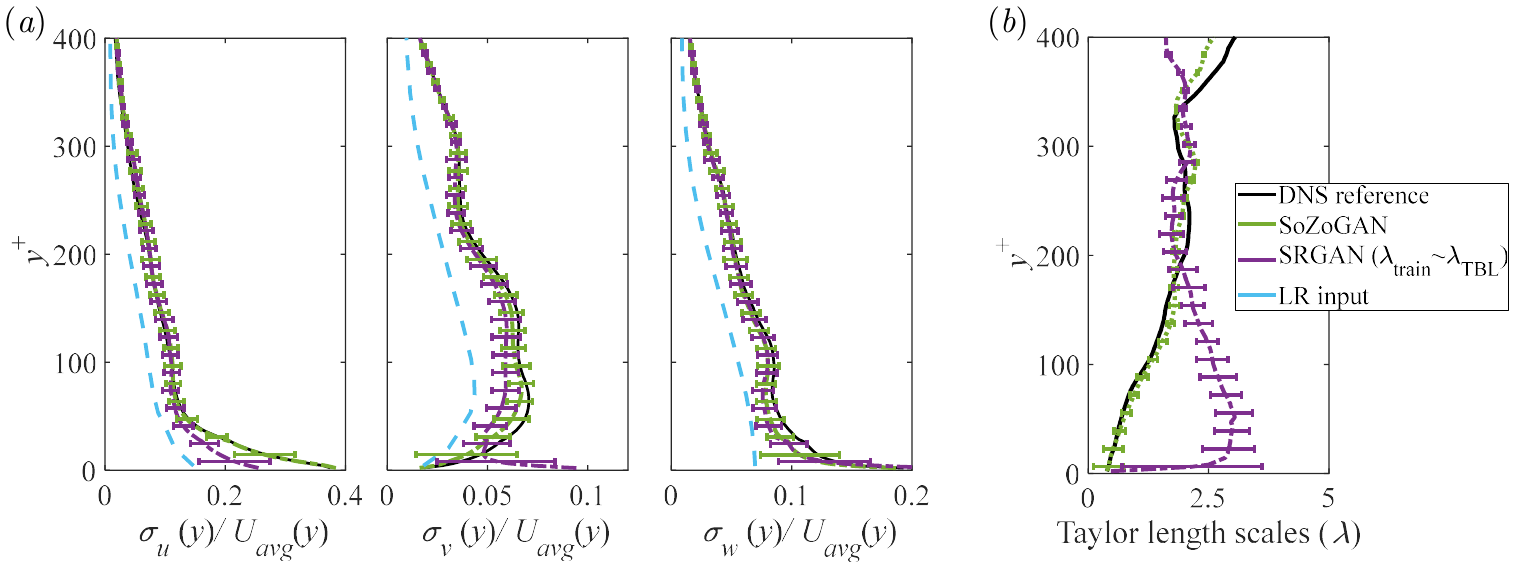}
\caption{Wall-normal profiles of (a) turbulent intensities and (b) Taylor length scales for global and zonal generation, both with microscale alignment; error bars show standard deviations at each wall distance.}\label{fig7jfm}
\end{figure}

To further assess the quantitative physical fidelity of the super-resolved fields, we examine and compare flow statistics between the generated and reference velocity fields. Figure~\ref{fig7jfm}(a) presents wall-normal profiles of the normalized turbulence intensities, $\sigma_u(y)/U_{\mathrm{avg}}(y)$, $\sigma_v(y)/U_{\mathrm{avg}}(y)$, and $\sigma_w(y)/U_{\mathrm{avg}}(y)$, obtained by averaging over 1000 snapshots. The proposed SoZoGAN framework closely reproduces the DNS reference profiles, accurately capturing the pronounced turbulence intensities near the wall that are absent in the coarse input fields. In contrast, the non-zonal baseline SRGAN presents substantial discrepancies across the viscous sublayer, buffer layer, and logarithmic region, failing to reconstruct the correct levels of turbulence intensity.

To provide a more direct assessment of microscale structure generation, we analyze the wall-normal profiles of the Taylor microscale length presented in figure~\ref{fig7jfm}(b). SoZoGAN, equipped with zonal generation, achieves excellent agreement with the DNS reference in the critical near-wall region ($y^+ \approx 0$-$150$), faithfully capturing the true eddy sizes. In contrast, the non-zonal baseline SRGAN systematically overestimates the Taylor microscales in this region. These findings underscore the effectiveness of SoZoGAN's zonal decomposition, which enables not only accurate reproduction of turbulent intensities but also precise recovery of average microscale eddy sizes, particularly in the most challenging near-wall portion of the wall-bounded turbulence.

\begin{figure}
\centering
\includegraphics[width=0.85\textwidth]{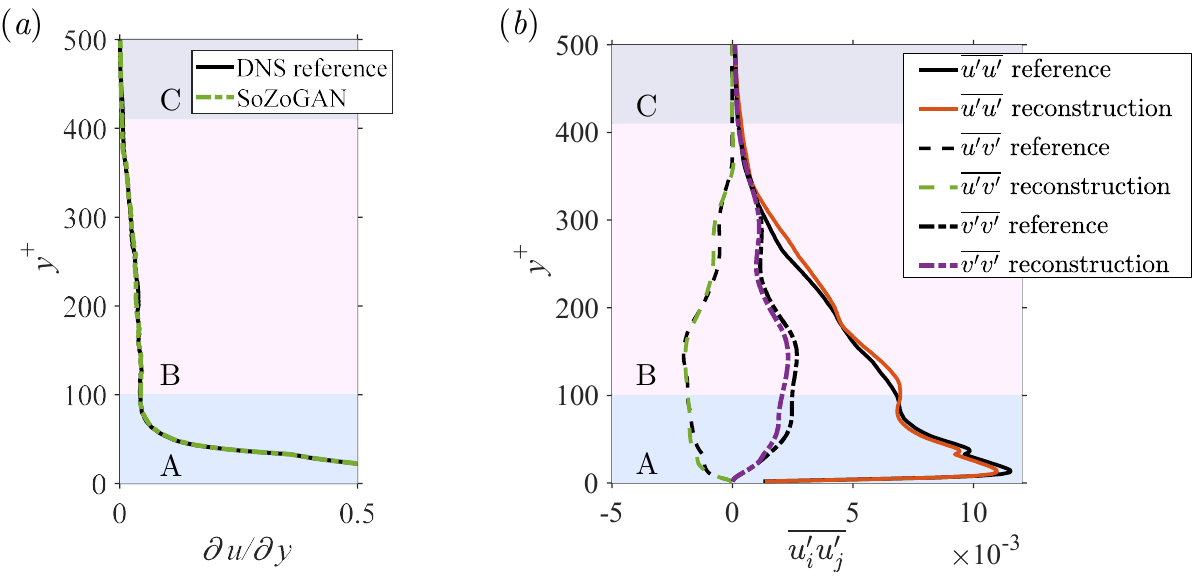}
\caption{Wall‑normal profiles of (a) vertical gradient of streamwise velocity, $\partial u / \partial y$, and (b) the Reynolds stress components, $\overline{u'u'}$, $\overline{u'v'}$, and $\overline{v'v'}$, spanning the wall‑normal range across all decomposed subdomains (labeled by A, B and C).}\label{fig7p5jfm}
\end{figure}

Furthermore, we examined the spatial continuity of the generated TBL field across subdomain interfaces by evaluating wall‑normal profiles of the streamwise‑velocity gradient $(\partial u/\partial y)$ (figure~\ref{fig7p5jfm}(a)) and the Reynolds‑stress components $\overline{u'u'}$, $\overline{u'v'}$, and $\overline{v'v'}$ (figure~\ref{fig7p5jfm}(b)). These profiles, covering the full wall‑normal range and all decomposed subdomains, agree closely with DNS and remain smooth across interface locations, with no discernible jumps or spurious fluctuations. Combined with the turbulence intensity and Taylor microscale distributions shown in figure~\ref{fig7jfm}, which likewise exhibit seamless behaviour at interfaces, these results confirm that SoZoGAN’s zonal generation strategy, together with the $1\times$ SRGAN smoothing step (figure~\ref{fig1}(c)), produces a globally continuous and physically consistent super‑resolved flow field.

\begin{figure}
\centering
\includegraphics[width=0.95\textwidth]{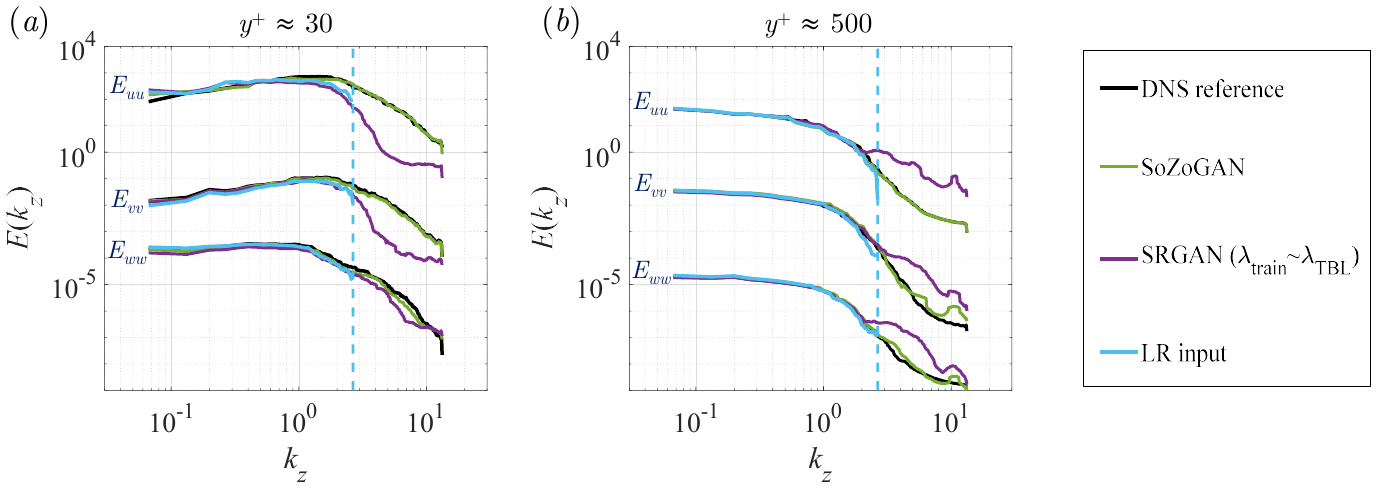}
\caption{Wavenumber spectra for the three velocity components, calculated at certain heights of (a) $y^+\approx30$ (near the wall) and (b) $y^+\approx500$ (faraway from the wall), comparing SoZoGAN and SRGAN. $E_{uu}$ is scaled by $1\times10^6$, $E_{vv}$ by $1\times10^3$, and $E_{ww}$ remains unscaled for clarity.}\label{fig8jfm}
\end{figure}

We finally investigate the turbulent energy distribution of the generated small-scale structures within the TBL to highlight SoZoGAN’s local spectral performance. Figure~\ref{fig8jfm} presents the horizontal wavenumber spectra at two representative wall-normal locations, computed over 1000 test snapshots. In the near-wall region ($y^+ \approx 30$), as shown in figure~\ref{fig8jfm}(a), SoZoGAN more accurately captures the energy cascade across the high‑wavenumber range, clearly outperforming the baseline SRGAN model pretrained with an average characteristic microscale ($\lambda_{\rm train}\approx \lambda_{\rm TBL}$). This advantage arises because the baseline model’s learned characteristic microscale tends to exceed the scales present near the wall. At higher wall-normal locations, closer to the free stream ($y^+ \approx 500$; figure~\ref{fig8jfm}(b)), SoZoGAN successfully avoids the overestimation of microscale turbulence energy. Such overprediction is a known limitation of the conventional sole-model approach, where the learned scale underestimates the actual eddy sizes farther from the wall. This comparison underscores SoZoGAN’s adaptive capacity to resolve spatially varying scales, significantly enhancing the spectral fidelity of the generated microscale turbulence throughout the distinct subregions of the TBL.

\subsubsection{Robustness with respect to subdomain partitioning and super-resolution ratio}\label{Robustness}

\begin{figure}
\centering
\includegraphics[width=1.0\textwidth]{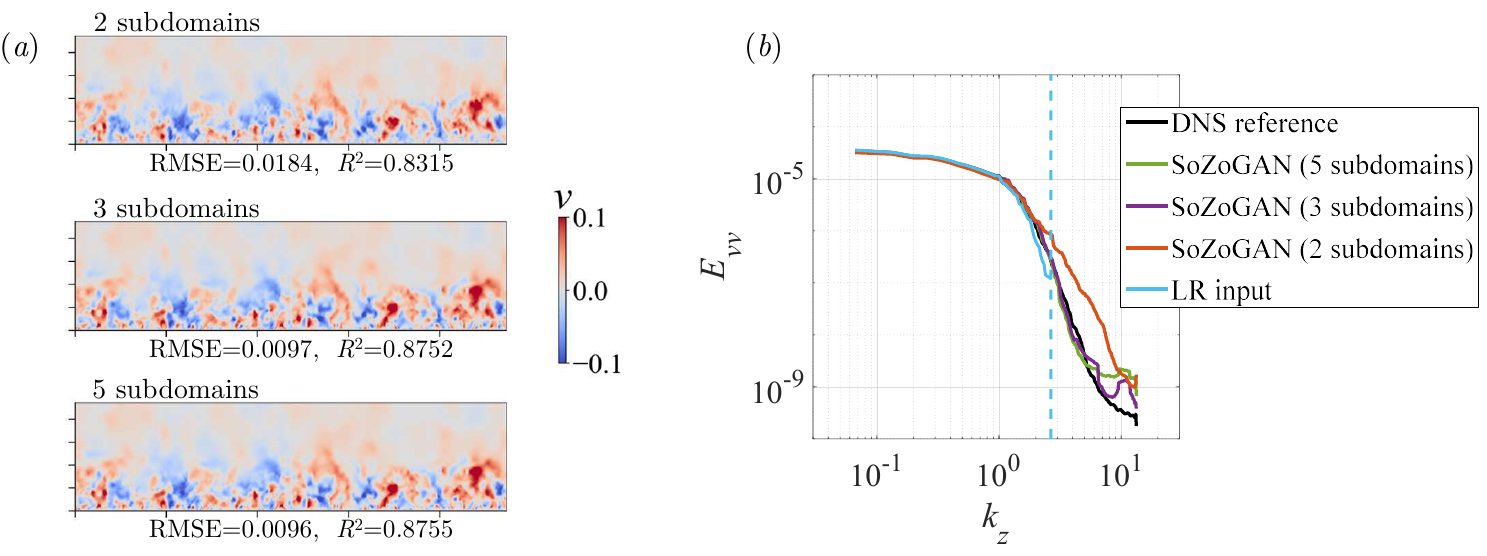}
\caption{Effect of the partitioning number on the SoZoGAN performance in TBL: (a) Synthesized instantaneous $v$ fields along with their corresponding (b) wavenumber spectra, using SoZoGAN based on 2, 3 and 5 partitioning subdomains.}\label{fig9jfm}
\end{figure}

\begin{figure}
\centering
\includegraphics[width=0.98\textwidth]{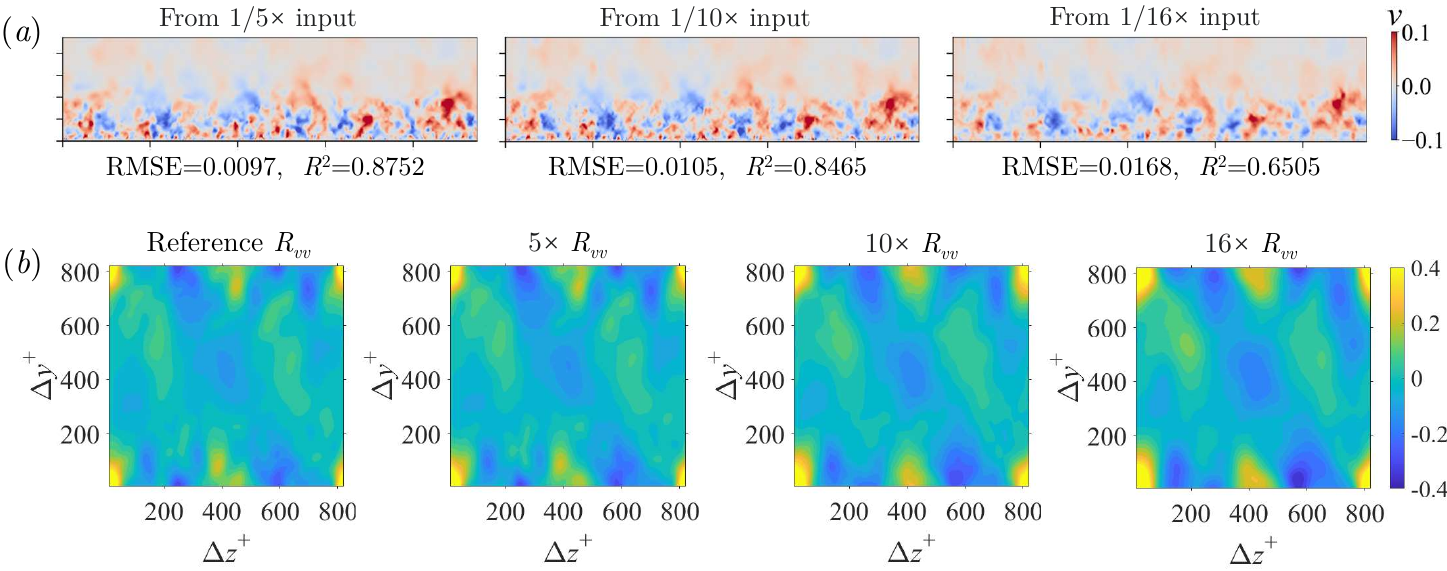}
\caption{Effect of the coarseness of the low-resolution inputs on the SoZoGAN performance in TBL: (a) Synthesized instantaneous $v$ fields and (b) cross-correlation coefficient contours of predicted snapshots from the $1/5\times$, $1/10\times$ and $1/16\times$ low-resolution inputs.}\label{fig10jfm}
\end{figure}

We investigate how the number of subdomains affects multi-scale turbulence generation accuracy and the framework's applicable super-resolution ratios, as shown in figure~\ref{fig9jfm} and figure~\ref{fig10jfm}, respectively.

Figure~\ref{fig9jfm} examines the impact of varying the number of decomposed subdomains on the accuracy of multi‑scale turbulence generation. TBL super-resolution using SoZoGAN are carried out with two, three, and five subdomains. As shown in figure~\ref{fig9jfm}(a), the instantaneous super‑resolved velocity fields produced with two subdomains exhibit larger generation errors than those with three subdomains. Increasing the partitioning from two to three subdomains reduces the RMSE by approximately 50\%, indicating a clear gain from finer zonal decomposition. By contrast, increasing the number from three to five subdomains produces only minor improvements in both the RMSE and $R^2$ metrics. 

The corresponding wavenumber spectra at $y^+ \approx 500$ are presented in figure~\ref{fig9jfm}(b). In the two‑subdomain case, the energy in the dissipation range is markedly overestimated relative to the DNS reference. This overestimation is substantially reduced when three subdomains are used, owing to a more accurate estimation of characteristic turbulent scales in the outer‑layer wake region. Improved scale estimation suppresses the introduction of spurious high‑frequency fluctuations during reconstruction. Further increasing the number of subdomains beyond three yields only modest spectral improvements while significantly increasing model complexity and computational cost.

Collectively, the three‑subdomain configuration achieves a strong balance between reconstruction accuracy and computational efficiency. It captures the wall‑normal variation of characteristic scales well enough to prevent spectral distortions, while avoiding the diminishing returns and extra cost in data preparation and model training that come with finer partitioning beyond three subdomains.

We further evaluate the applicability of the proposed SoZoGAN framework for TBL generation across a range of super-resolution ratios ($r$). As illustrated in figure~\ref{fig10jfm}(a), instantaneous fields are reconstructed from $1/5	\times$, $1/10 \times$, and $1/16 \times$ low-resolution inputs using $r \times$ SoZoGAN ($r=[5,10,16]$). For $r = 5$ and $10$, SoZoGAN effectively reproduces the near-wall microscale vortex structures, maintaining consistently low RMSE values and near-unity $R^2$ values, indicative of excellent agreement with DNS references. When the ratio increases to $r = 16$, SoZoGAN still captures the major flow patterns present in the DNS fields, despite the much coarser input resolution. However, the recovery of near-wall microscale structures diminishes, both in terms of their number and intensity. This limitation is reflected in a marked increase in RMSE and a noticeable decline in $R^2$, underscoring the greater challenge of recovering fine-scale features at higher super-resolution ratios.

To probe the spatial structures of the SoZoGAN-generated fields, figure~\ref{fig10jfm}(b) presents the spatial cross-correlation coefficients $R_{vv}(\Delta y^+,\Delta z^+)$ for wall-normal velocity at different super-resolution ratios. Across all ratios examined, the reconstructed fields yield correlation patterns largely consistent with DNS. Nevertheless, as $r$ increases, the absolute values of these coefficients tend to be slightly overestimated, reflecting a loss of spatial complexity due to the underrepresentation of microscale structures in the generated fields.

Overall, these results confirm the robustness and accuracy of SoZoGAN in generating microscale turbulence for super-resolution ratios up to $r=10$. Even at $r=16$, the framework preserves the fundamental spatial characteristics of high-resolution wall turbulence, albeit with some loss of detail. Beyond this ratio, individual SRGAN models within the SoZoGAN framework reach their capacity limits of super-resolution, and increasing the number of subdomains alone does not yield further improvements in predictive accuracy.

\subsection{Inhomogeneous turbulence: turbulent channel flow}

The next application focuses on the super-resolution generation of a turbulent channel flow with a friction Reynolds number of $\mathrm Re_{\tau}= 1000$ \citep{bib27}. Similar to the TBL, the turbulent channel flow is wall-bounded and exhibits significant scale variation in the wall-normal direction. However, there are important distinctions between the two types of turbulence.

In the outer region (far from the wall) of the TBL, turbulence tends to become smoother and less intense, with velocity fluctuations weakening as the distance from the wall increases. In contrast, turbulence in the channel flow remains relatively active and exhibits significant spatial fluctuations even far from the walls. This is primarily due to the confinement imposed by the two parallel walls, which restricts the dissipation of turbulent energy. As a result, super-resolving turbulent channel flow presents a unique challenge.

\subsubsection{``Zero-shot'' small-scale generation using SoZoGAN}

For this test case, the velocity data of the turbulent channel flow are also obtained from a DNS database provided by JHTDB. The coordinate system and corresponding velocity components adhere to the definitions used in the TBL database. To focus on the generation of wall turbulence, the data are extracted from a $y-z$ section near one of the channel walls. The section is $y\times z$ = 0.3835$H$ × 1.5340$H$ with $500 \times 500$ evenly distributed grid nodes, where $H$ is the channel height. To obtain the low-resolution inputs, the original DNS fields are downsampled to $50\times 50$ using a factor of 10. A total of 1000 snapshots are selected, evenly distributed across the entire duration of the simulation, ensuring a broad statistical representation of the turbulence.

\begin{figure}
\centering
\includegraphics[width=0.9\textwidth]{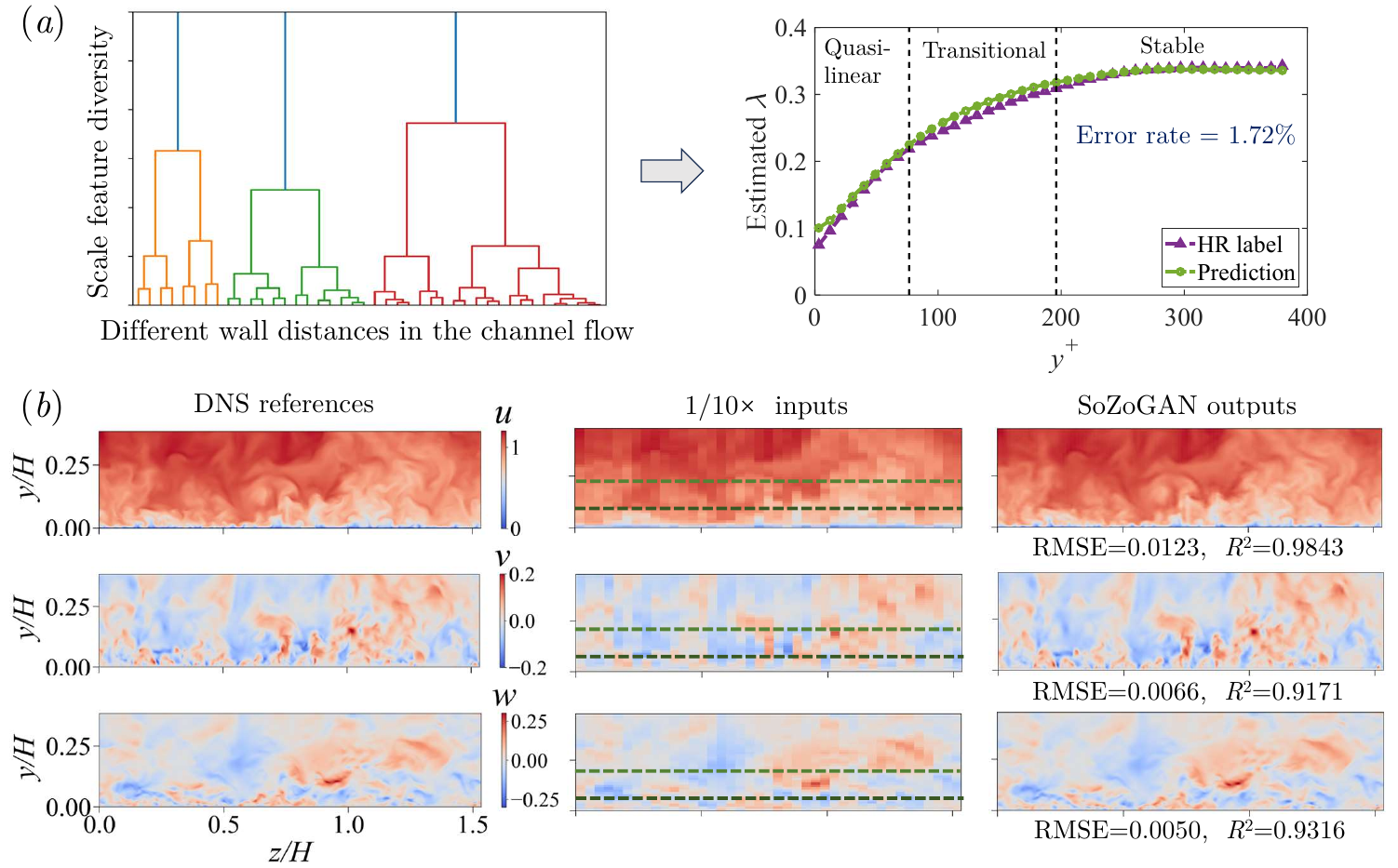}
\caption{``Zero-shot'' generation of vertical $y$-$z$ plane in turbulent channel flow by the proposed SoZoGAN. (a) Zonal decomposition and microscale estimation of the turbulent channel flow. Three scale‑related subdomains are decomposed. Taylor microscale estimation is performed along the wall-normal direction of the turbulent channel flow. (b) Generated instantaneous fields of three velocity components with their reconstruction errors (RMSE) and accuracies ($R^2$). The green dash lines shown on the low-resolution inputs (middle panels) represent the subdomain boundaries identified using hierarchical clustering.}\label{fig11jfm}
\end{figure}

Subdomains are identified using hierarchical clustering based on the wavenumber spectra of low-resolution inputs. The corresponding estimated Taylor microscales, obtained via a MLP, are shown in figure~\ref{fig11jfm}(a). Owing to the similar generation mechanisms governing wall-bounded turbulence in both the TBL and channel flow, the same procedures for zonal decomposition and microscale estimation are employed. These procedures effectively capture both the magnitude and the wall-normal variation of microscales in the DNS reference, resulting in a low spatially averaged error ($1.72\%$). Across the channel, three distinct scale‑related subdomains emerge: quasi‑linear, transitional, and stable subdomains.

After zonal decomposition and microscale estimation, we perform the scale-oriented generation of small-scale channel flow. As depicted in figure~\ref{fig11jfm}(b), the synthesized $u$, $v$ and $w$ velocity fields exhibit agreement with the DNS results. The large spatial fluctuations away from the wall are well replicated through the ``zero-shot'' transfer of the SoZoGAN framework. The RMSE errors are low and the $R^2$ accuracies all exceed 0.9, indicating that the SoZoGAN generation accurately captures the multi-scale turbulent structures.

\begin{figure}
\centering
\includegraphics[width=1.0\textwidth]{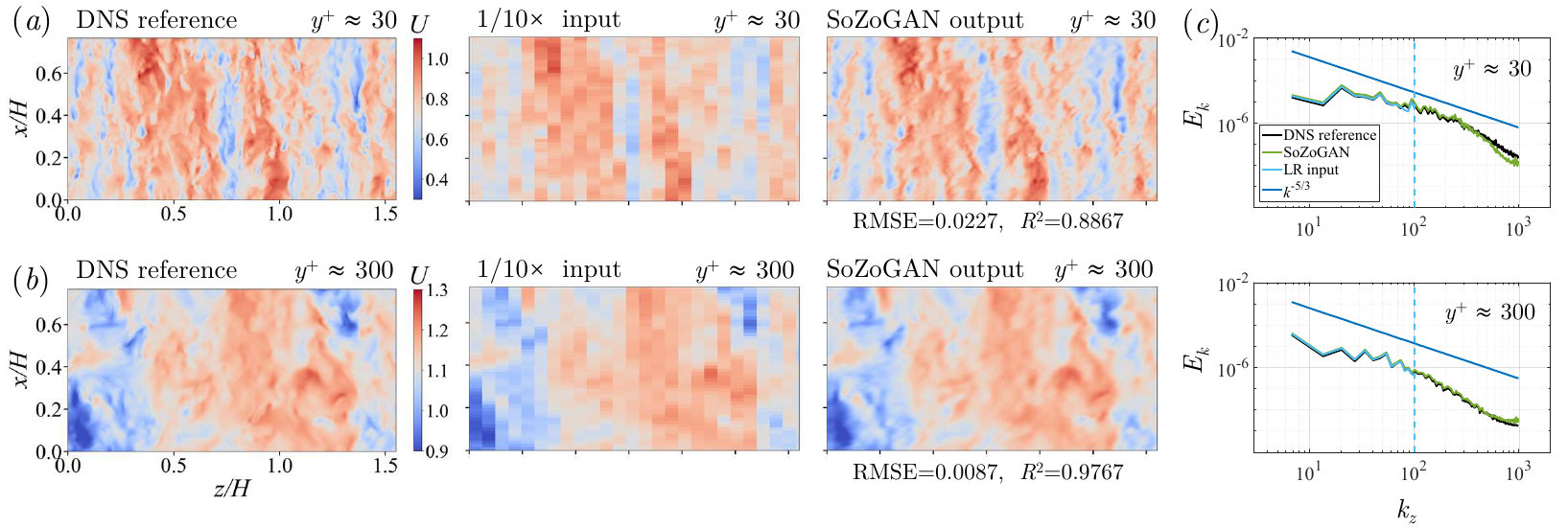}
\caption{``Zero-shot'' generation of horizontal $x$-$z$ planes in turbulent channel flow by the proposed SoZoGAN. Instantaneous total velocity $U$ fields at two wall distances, (a) $y^+ \approx 30$ (near-wall region) and (b) $y^+ \approx 300$ (outer-layer region), reconstructed at 10$\times$ super-resolution from low-resolution inputs. Their reconstruction errors (RMSE) and accuracies ($R^2$) are marked. (c) Wavenumber spectra of the reconstructed $U$ distributions for both wall distances.}\label{fig11p5jfm}
\end{figure}

To further assess the generality and Galilean invariance of SoZoGAN, we conducted additional tests on horizontal $x$-$z$ planes of channel flow at two  wall distances: $y^+ \approx 30$ (near-wall region) and $y^+ \approx 300$ (outer-layer region). These planes were extracted from the same DNS database. Figure~\ref{fig11p5jfm}(a) and (b) presents the instantaneous total velocity $U$ fields generated in a zero-shot manner, while figure~\ref{fig11p5jfm}(c) shows the corresponding wavenumber spectra for both wall distances. In the near-wall case (figure~\ref{fig11p5jfm}(a)), which is characterized by pronounced streak-like structures, SoZoGAN faithfully reproduces the fine-scale velocity fluctuations observed in the DNS reference, achieving $R^2$ values close to 0.9 despite the complexity of the flow. In the outer-layer case (figure~\ref{fig11p5jfm}(b)), where the turbulent structures are significantly larger in scale and more isotropic, the generated fields exhibit similarly high agreement with the DNS data and preserve the spatial organization of coherent motions.

The microscale alignment in SoZoGAN plays a key role in distinguishing scale-specific features at different wall distances. Smaller near-wall scales and larger outer-layer structures are both accurately recovered without any additional training. The wavenumber spectra further confirm that SoZoGAN reproduces the inertial-subrange energy characteristics appropriate to each wall distance (figure~\ref{fig11p5jfm}(c)). Consistent reconstruction performance across both $x$-$y$ and $x$-$z$ cross-planes, and at wall distances with notably different turbulent scales, provides strong evidence that the SoZoGAN satisfies Galilean invariance. This invariance implies that the same pretrained generative models can be applied in uniformly translating reference frames without loss of reconstruction fidelity.

\subsubsection{``Zero‑shot'' performance of SoZoGAN versus other state‑of‑the‑art models}

We examine the performance of SoZoGAN in comparison with several state-of-the-art super-resolution models for microscale turbulence generation in channel flow. The baseline models include hDSC-MS, a deterministic approach that incorporates multiscale modeling to capture both large and small-scale turbulent structures \citep{bib30, bib39}. In addition, two generative models are considered: Diffusion \citep{bib38}, which employs a progressive denoising framework to iteratively refine coarse inputs by learning the conditional score function of turbulence statistics, and SRGAN, which serves as the base super-resolution model embedded within SoZoGAN. Unlike SoZoGAN, all baseline models do not incorporate zonal decomposition or scale alignment. Instead, they are trained on original HIT samples ($\lambda_1' \sim 0.0331$) and directly transferred, without fine-tuning, to the channel flow scenario for small-scale generation.

\begin{figure}
\centering
\includegraphics[width=0.97\textwidth]{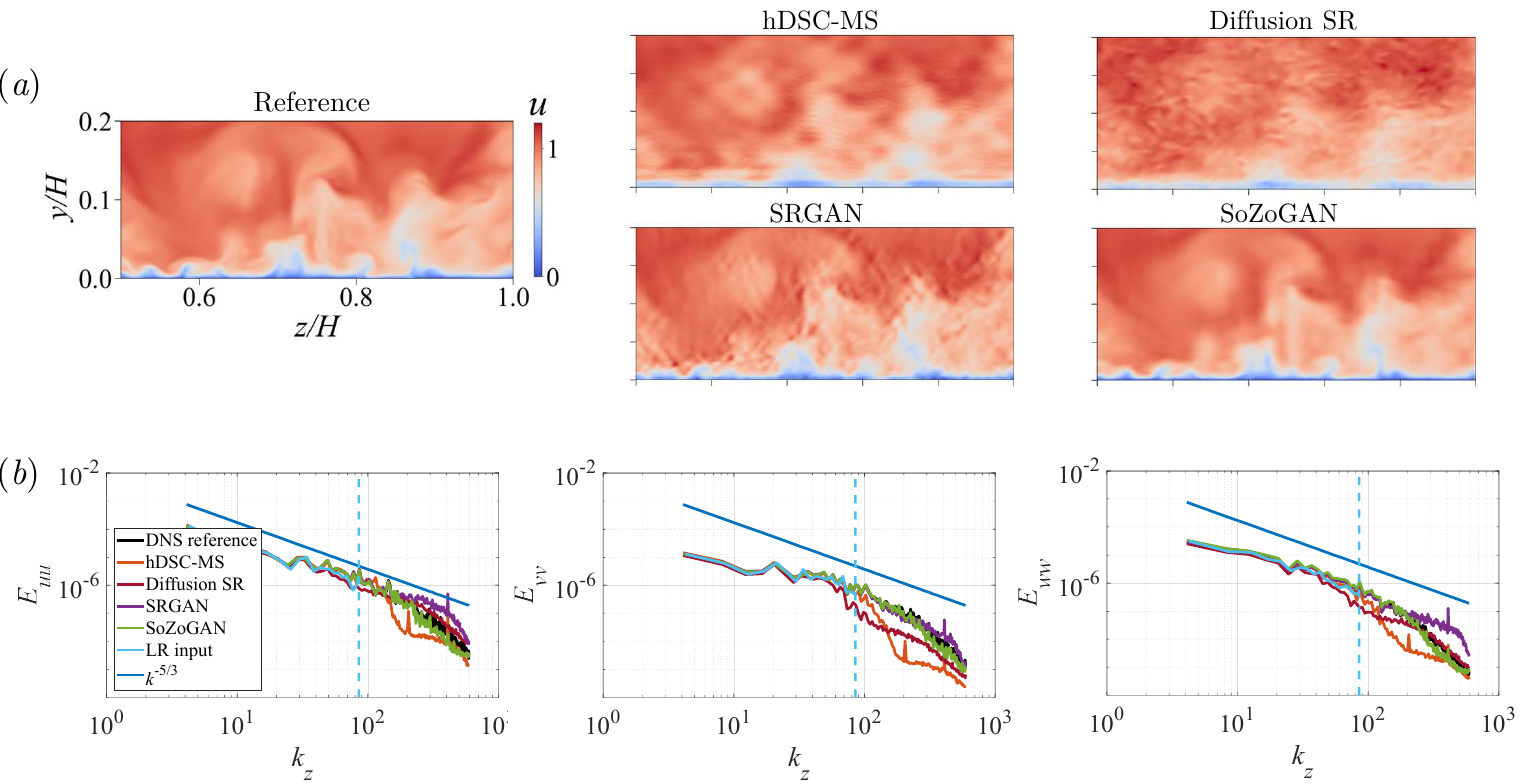}
\caption{Comparison of ``zero-shot’’ generation for microscale turbulent channel flow using different super-resolution models: (a) Instantaneous streamwise velocity fields and (b) wavenumber spectra of three velocity components averaged over the region of $y^+<400$ predicted by hDSC-MS, Diffusion, SRGAN, and SoZoGAN, compared with the DNS reference.}\label{fig12jfm}
\end{figure}

Figure~\ref{fig12jfm}(a) presents a comparative analysis of instantaneous streamwise velocity fields generated by these models. The hDSC-MS model fails to resolve many critical fine-scale features, resulting in excessively smoothed fields that lack the distinctive small-scale vortical details characteristic of near-wall turbulence. Meanwhile, the generative Diffusion and SRGAN models introduce spurious, non-physical fluctuations - a consequence of inadequate calibration to local turbulent scales and limited adaptability to the heterogeneity of wall-bounded flows. In sharp distinction, SoZoGAN faithfully reconstructs the complex scale variation observed across the wall-normal direction in channel flow. It accurately recovers intricate small-scale roll-up structures near the wall, while smoothly transitioning to larger-scale features farther from the wall, yielding instantaneous velocity fields in close agreement with DNS.

The superiority of SoZoGAN is expected to be underscored in the wavenumber spectra comparison. In figure~\ref{fig12jfm}(b), the wavenumber spectra of three velocity components averaged over the region of $y^+<400$. The spectra produced by SoZoGAN most closely match DNS results across the entire range of wavenumbers. hDSC-MS underestimates turbulence energy at small scales, reflecting its limited microscale reconstruction capability. While Diffusion achieves reasonable small-scale energy recovery, it fails to preserve energy in the inertial subrange, indicating inconsistent multi-scale turbulence generation. SRGAN, on the other hand, maintains reasonable accuracy in the inertial subregion but tends to overpredict turbulence energy at the finest scales. Overall, these results underscore that the scale-oriented zonal decomposition and microscale alignment strategies of SoZoGAN provide critical advantages for robust and accurate microscale turbulence generation in wall-bounded flows.

Despite the need to pretrain multiple specialized SRGANs, the total training cost of SoZoGAN remains moderate (about 18 GPU-hours on two NVIDIA Tesla V100 GPUs) and is a one‑time offline investment. Inference of SoZoGAN is somewhat slower than the baselines due to the zonal generation and subdomain merging procedures; however, for the channel flow case reported here, the average inference time per snapshot is only about 0.075 seconds on the same hardware. These computational demands are well within practical limits and are more than offset by the gains in multi‑scale generation accuracy and robustness.

\subsection{Evaluations of zero-shot transfer in SoZoGAN}

To emphasize the effectiveness of the proposed zero‑shot transfer strategy, we introduce an ``in‑domain model'' baseline. This baseline uses an SRGAN with exactly the same network architecture, loss function, and hyperparameter settings as those SRGANs employed in SoZoGAN. The key difference lies in the training setup:  
(i) ``in‑domain model'' — trained and tested entirely within the same flow type using a standard train–test split from its DNS dataset;  
(ii) ``cross‑domain model'' — operates in a zero‑shot, cross‑domain setting, with each SRGAN pretrained once on a single HIT dataset and applied without retraining or fine‑tuning to generate SR fields for completely unseen target flows, including HIT at different Reynolds numbers and spatially inhomogeneous turbulence.  
This comparison contrasts an idealized best‑case scenario with abundant target‑flow data (``in‑domain model'') against the more challenging and practically relevant zero‑shot inference task (``cross‑domain model'') addressed by SoZoGAN.

\begin{figure}
\centering
\includegraphics[width=1.0\textwidth]{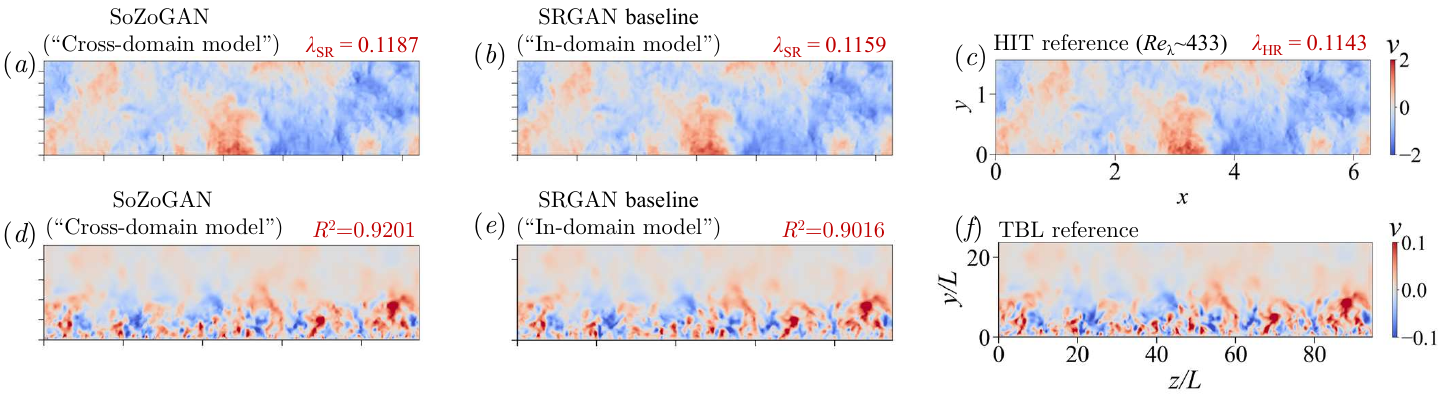}
\caption{Effectiveness of zero-shot transfer in SoZoGAN. (a)-(c) Small-scale HIT generations: instantaneous $v$-component fields from (a) SoZoGAN (``cross‑domain model''), (b) SRGAN baseline (``in‑domain model'')  trained directly on target HIT data, and (c) DNS reference. Reported $\lambda_{\text{SR}}$ and $\lambda_{\text{HR}}$ indicate the Taylor microscales associated with reconstructed and DNS fields. (d)-(f) Small-scale TBL generations: instantaneous $v$-component fields from (d) SoZoGAN, (e) SRGAN baseline trained directly on target TBL data, and (f) DNS reference. The $R^2$ value shown over each generated field is averaged across the generated $u$, $v$, and $w$ fields.}\label{fig13jfm}
\end{figure}

Figure~\ref{fig13jfm} compares turbulence‑generation results for both HIT and TBL cases. In HIT, the instantaneous $v$‑component fields produced by SoZoGAN (figure~\ref{fig13jfm}(a)) are visually and statistically indistinguishable from both the DNS reference (figure~\ref{fig13jfm}(c)) and the ``in‑domain model'' output (figure~\ref{fig13jfm}(b)). The Taylor microscales $\lambda_{\text{SR}}$ and $\lambda_{\text{HR}}$ derived from SoZoGAN, the ``in‑domain model'', and DNS data exhibit excellent agreement, confirming the accurate reproduction of key small‑scale structures in isotropic turbulence. In TBL, SoZoGAN likewise generates instantaneous structures in close agreement with both DNS (figure~\ref{fig13jfm}(f)) and the ``in‑domain model'' (figure~\ref{fig13jfm}(e)), while its scale‑oriented zonal generation strategy captures the wall‑normal variation of characteristic turbulent scales more faithfully. This improvement results in a modest increase in $R^2$ accuracy over the ``in‑domain model'', underscoring the benefit of localized scale‑aware generation in wall‑bounded flows.

Overall, these results demonstrate that even in the zero‑shot ``cross‑domain model'' setting, SoZoGAN achieves turbulence‑generation accuracy comparable to the ``in‑domain model'' trained directly on target‑flow data. This confirms the effectiveness of the proposed method and demonstrates that high fidelity turbulence generations are possible with substantially reduced data requirements. Operating without any high resolution target flow training data makes SoZoGAN particularly valuable in scenarios where such data are scarce or unavailable, while still preserving the physical consistency of the reconstructed fields.

\subsection{Discussions on mechanisms of SoZoGAN generalization}

The reason for this robust generalization finds its explanation in the foundations of turbulence theory. SoZoGAN can transfer effectively across very different turbulence regimes, thanks in large part to Kolmogorov's local isotropy hypothesis \citep{K41}. This hypothesis suggests that the small-scale statistics of turbulence are nearly universal and largely independent of the large-scale anisotropy of the flow. Empirical evidence from previous spectral analyses and second-order statistical studies \citep{Saddoughi1994, Shen2000} supports the idea that small-scale turbulence features are sufficiently generic. This means that models pretrained on HIT can be reasonably extended to more complex, strongly sheared flows. In addition, SoZoGAN’s loss function incorporates a physical constraint—the residual of the incompressible continuity equation—which further aids adaptation to flows with pronounced anisotropy and strong shear. Our results confirm that SoZoGAN sustains high accuracy, not only for velocity spectra but also for higher-order velocity moments (up to fourth order) over a wide range of turbulence scenarios.

The model’s strong performance in representing the statistics of small-scale turbulent motions is evident in its agreement with DNS for both velocity fields and low-order velocity gradient moments (see table \ref{tab2}). For many engineering and geophysical applications, this level of accuracy is sufficient. Nevertheless, persistent small-scale anisotropy—especially in wall-bounded flows with prominent mean shear \citep{Shen2000}—remains a challenge for transfer learning based solely on HIT data. In these cases, third and fourth-order velocity gradient moments show an average error of about 18\%. This highlights the inherent limitation of capturing subtle features of small-scale anisotropy using HIT-pretrained models. Future studies might overcome this limitation by explicitly modeling the effects of shear and intermittency \citep{Shen2000, Buaria2025}, with special attention to the fine-scale moments that are critical in near-wall regions for momentum and scalar transport.

\subsection{Adjustment of SoZoGAN to adapt to energy-filtered coarse inputs} \label{subsec37}

The scale-oriented zonal framework proposed in this study is architecture‑agnostic, and can readily extend beyond GANs (SoZoGAN) to other deep learning SR models. This flexibility further broadens its applicability to diverse turbulence reconstruction scenarios. In particular, in practical numerical simulations with extremely coarse grids, the resulting LR data may exhibit premature or excessive attenuation of turbulent kinetic energy due to strong numerical dissipation. Such cases deviate from the moderate downsampling regime considered for the SoZoGAN discussed in previous sections, where the LR inputs are generated through well‑controlled average pooling that preserves energy in the energy‑containing motions and part of the inertial subrange.

To address these highly energy‑filtered LR inputs, the SRGAN‑based energy‑cascade model (EC‑SRGAN) proposed by \citet{bib16} can be embedded within the present scale‑oriented zonal framework, yielding a hybrid model referred to here as SoZoEC‑SRGAN (a variant of SoZoGAN). In the approach of \citet{bib16}, the LR fields are first obtained by average‑pooling from DNS turbulent fields, followed by a low‑pass filtering step. This additional filtering mimics the enhanced energy dissipation in the energy‑containing range and inertial subrange observed in very coarse-grid simulations. In our hybrid integration, the GAN component of SoZoGAN is replaced by EC‑SRGAN, whose architecture and loss functions were designed specifically to compensate for such energy deficits and recover fine‑scale turbulent motions in wall‑bounded flows from strongly energy‑filtered inputs.

\begin{figure}
\centering
\includegraphics[width=1.0\textwidth]{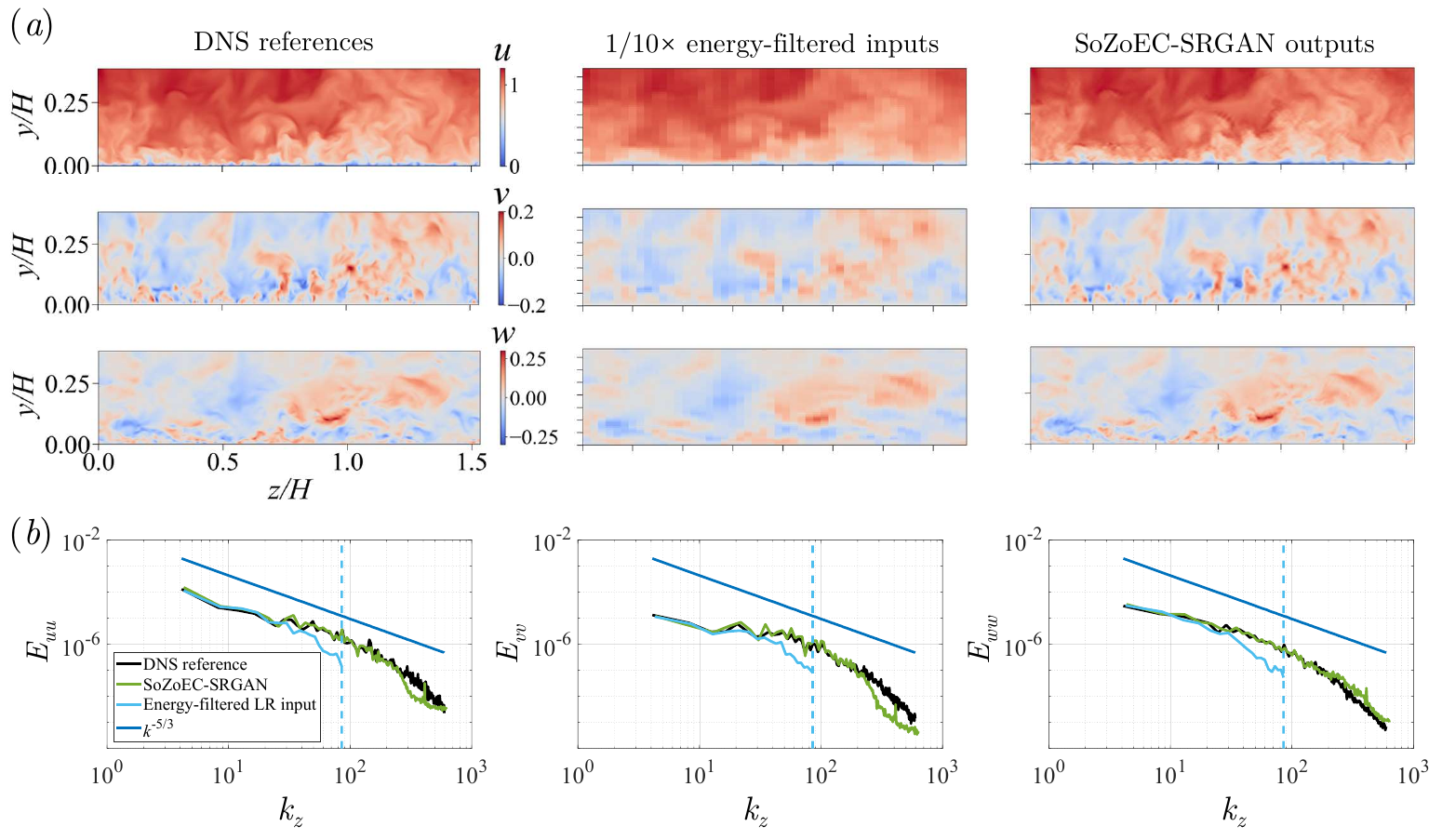}
\caption{``Zero-shot'' generation of turbulent channel flow from strongly energy filtered LR inputs using the proposed SoZoEC-SRGAN. (a) Instantaneous fields of the three velocity components, comparing LR inputs obtained by average pooling and subsequent low pass filtering of DNS data, $10\times$ super-resolved outputs from SoZoEC-SRGAN, and the DNS reference fields. (b) Corresponding wavenumber spectra of three velocity components for the LR inputs, super-resolved outputs, and DNS references.}\label{fig14jfm}
\end{figure}

As a representative demonstration, we apply SoZoEC‑SRGAN to a turbulent channel flow case with LR input fields exhibiting severe energy loss, corresponding to a $10\times$ SR task. Figure~\ref{fig14jfm}(a) compares instantaneous fields of the three velocity components from the energy‑filtered LR inputs, the SoZoEC‑SRGAN reconstructions, and the high‑resolution DNS references. The SoZoEC‑SRGAN outputs successfully recover spatially coherent near‑wall fine‑scale structures that are largely absent in the LR input, with flow features and spatial distributions closely matching those in the DNS fields. Figure~\ref{fig14jfm}(b) presents the corresponding wavenumber spectra, showing that SoZoEC‑SRGAN effectively restores energy in the lower part of the inertial subrange and extends the $-5/3$ energy cascade towards finer scales. This indicates that SoZoEC‑SRGAN can reconstruct statistically accurate small‑scale turbulence contents despite substantial energy deficits in the coarse inputs.

This example illustrates that the proposed scale‑oriented zonal framework can adaptively incorporate SR models with complementary strengths to target specific deficiencies in the input data. In particular, by selecting or integrating architectures capable of compensating for energy loss, the framework extends the applicability of zero‑shot turbulence super‑resolution to extreme coarse‑grid scenarios in numerical settings.

\subsection{Discussions on limitations of SoZoGAN}\label{limitations}

\begin{figure}
\centering
\includegraphics[width=1.0\textwidth]{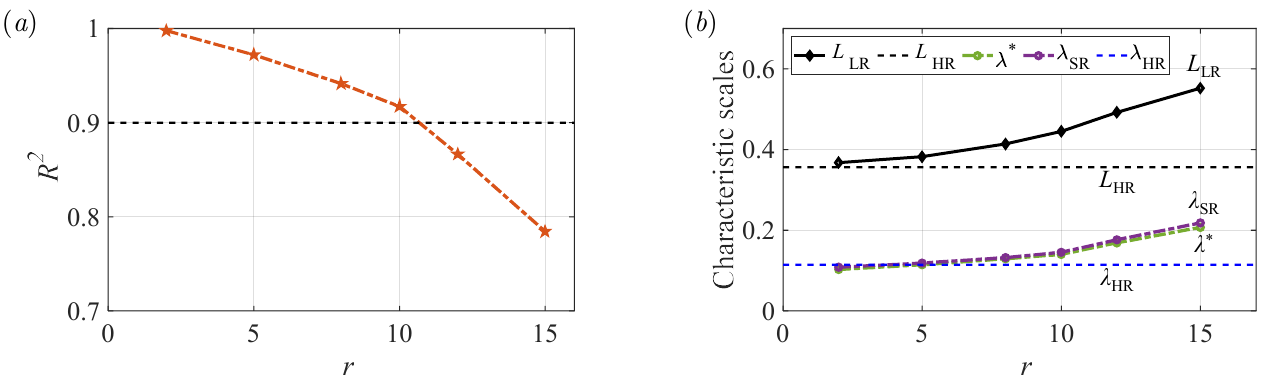}
\caption{Robustness of turbulence super-resolution to input coarsening with SoZoGAN. (a) Accuracy of the super-resolved velocity fields, quantified by the coefficient of determination ($R^2$), as a function of the downsampling factor. (b) Predicted Taylor microscales of the SoZoGAN output ($\lambda_{\rm SR}$), MLP-estimated Taylor microscales from coarse inputs ($\lambda^*$) and corresponding input integral scales ($L_{\rm LR}$) at varying levels of input coarsening. The blue and black dash lines represent the calculated Taylor and integral scales based on the high-resolution reference fields, respectively. These results reveal the critical threshold in input resolution required for reliable small-scale reconstruction.}\label{fig4jfm}
\end{figure}

The effectiveness of turbulence super-resolution with SoZoGAN is fundamentally governed by the relationship between input grid resolution (the smallest resolvable turbulence scale in input data) and the characteristic length scales of turbulence. A key challenge lies in ensuring that the coarse input grid is sufficiently fine to retain the information needed for accurate recovery of small-scale structures. Through systematic analysis of homogeneous turbulence reconstruction (summarized in figure~\ref{fig4jfm}), we identify a distinct resolution threshold that determines the feasibility of super-resolution.

Figure~\ref{fig4jfm}(a) quantifies SoZoGAN performance across varying degrees of input coarsening using the coefficient of determination:
\begin{equation}
R^2 = \frac{1}{n} \sum_{j=1}^{n} \left[ 1 - \frac{\sum_{i=1}^{m}(y_{{\rm HR},i} - y_{{\rm SR},i})^2}{\sum_{i=1}^{m}(y_{{\rm HR},i} - \bar{y}_i)^2} \right],
\end{equation}
where $y_{\rm HR,\it i}$ is the DNS reference velocity at the $i$th grid point, $y_{\rm SR,\it i}$ is the predicted velocity, $\bar{y}_i$ is the DNS mean at that grid point, and $m$, $n$ denote the number of grid points and snapshots. Note that $\lambda_{\rm HR}$ represents the Taylor microscale calculated from the DNS reference field. When the coarse grid spacing is below approximately $1.1\lambda_{\rm HR}$ ($1/8	\times$ low-resolution input), SoZoGAN achieves high predictive accuracy ($R^2 > 0.9$), producing small-scale turbulence that is both visually and physically credible. However, as the grid spacing increases beyond $1.4\lambda_{\rm HR}$ ($1/10 \times$ low-resolution input), the ability to recover fine-scale features diminishes rapidly. This deterioration in performance is closely tied to the coarse input’s capacity to retain large-scale flow features and support precise estimation of the microscale, as shown in figure~\ref{fig4jfm}(b). For input grids no coarser than $1.1\lambda_{\rm HR}$, the integral scale $L_{\rm LR}$ deviates from the DNS benchmark $L_{\rm HR}$ by less than $16\%$, and the MLP predicts the microscale $\lambda^*$ with an average error of only $7.8\%$. In this regime, SoZoGAN reconstructs fields whose microscale $\lambda_{\rm SR}$ is close to the DNS reference with just $8.2\%$ error, indicating acceptable physical consistency. When the grid spacing exceeds $1.4\lambda_{\rm HR}$, however, $L_{\rm LR}$ deviates by more than $25\%$ and $\lambda^*$ is substantially overestimated, producing SoZoGAN-generated fields with $\lambda_{\rm SR}$ more than $27\%$ larger than the DNS value. Such deviations significantly undermine the physical fidelity of the reconstructed turbulence.

The interplay between input resolution and recoverable turbulence scales suggests a practical strategy for computational fluid dynamics. Suppose one wishes to generate turbulence with a target Taylor microscale $\lambda_{\rm HR}$. In this case, a rapid numerical simulation can be performed using a grid with a spacing of about $1.1 \lambda_{\rm HR}$ to produce a coarse flow field. This field then serves as input to SoZoGAN, which reconstructs the corresponding fine-scale turbulent features both efficiently and accurately. Such an approach greatly accelerates turbulence generation and offers a promising new pathway for CFD simulations.

Our findings also crystallize two main guidelines for applying the SoZoGAN framework. First, as discussed above, it is crucial to preserve information about the macroscale structures in the coarse input. If the grid becomes too coarse to resolve these structures, accurate prediction of small-scale turbulence is no longer feasible. 

Second, the Taylor-scale Reynolds number $\mathrm{Re}_\lambda$ present in the pretraining HIT dataset should reach at least as large as those in the testing turbulence. The flip between the training and testing $\mathrm{Re}_\lambda$ might lead to incorrect generated turbulence. Recent work \citep{bib31,Yeo2024} confirms that models trained at lower~$\mathrm{Re}_\lambda$ fail to reconstruct the small-scale features of higher~$\mathrm{Re}_\lambda$ flows. This is because higher~$\mathrm{Re}_\lambda$ turbulence exhibits broader distributions of velocity gradient invariants (Q-R space) than lower-$\mathrm{Re}_\lambda$ data can capture. Consequently, models trained on lower~$\mathrm{Re}_\lambda$ must extrapolate beyond their training range when applied to higher-$\mathrm{Re}_\lambda$ flows, resulting in underprediction of scale-invariant turbulent features. In this work, high-to-low $\mathrm{Re}_\lambda$ transfer is achieved by scaling up the original HIT field to create training samples with larger characteristic scales, improving compatibility with lower-$\mathrm{Re}_\lambda$ super-resolution targets. Conversely, a ``scaling-down'' approach could be used to emulate scale-invariant features characteristic of high-$\mathrm{Re}_\lambda$ flows, enriching low-$\mathrm{Re}_\lambda$ datasets and potentially enhancing low-to-high $\mathrm{Re}_\lambda$ generalization. This will be explored in future work.

Finally, although SoZoGAN in this paper employs subdomain decomposition only along the wall-normal direction, the method is readily extendable to more localized partitioning, as demonstrated in figure~\ref{fig2jfm}(b). While the current implementation uses regular rectangular subregions for convolutional processing, it can be straightforwardly generalized to adaptively defined subdoamins that more effectively capture the complex spatial variations present in real turbulent flows \citep{bib10}. By selecting appropriate partitioning strategies, the SoZoGAN framework is flexible enough to accommodate a wide range of flow inhomogeneities and geometric configurations.

\section{Conclusions}\label{Conclusions}

This study proposes SoZoGAN, a novel framework for efficient, high-fidelity turbulence super-resolution that generalizes across diverse flow types using only coarse-grained inputs. Unlike conventional data-intensive methods, SoZoGAN leverages pretraining on a single, readily available HIT dataset. Through strategically designed scaling transformations of HIT data, the framework constructs a library of scale-specific SRGAN models, enabling comprehensive coverage of turbulent scales.

A central feature of SoZoGAN is its capability to exploit the universality of small-scale turbulence dynamics. This allows for robust ``zero-shot’’ generation of turbulence across previously unseen flow scenarios, including not only homogeneous isotropic turbulence (HIT) but also boundary layer and channel flow distinct from the training data. Another key innovation is the zonal decomposition strategy, which leverages local, scale-sensitive physical quantities to guide the partitioning of input fields. This approach addresses the inherent spatial heterogeneity of turbulent flows and enhances the model’s adaptability to complex, non-homogenous turbulent structures. Compared with existing state-of-the-art turbulence super-resolution models, SoZoGAN demonstrates distinct advantages in two crucial aspects: the physical fidelity of small-scale turbulence reconstruction and adaptability to spatially heterogeneous flows. Moreover, the scale-oriented and architecture-agnostic design of SoZoGAN ensures compatibility with a wide range of deep learning models beyond GAN-based networks, offering the potential for further methodological innovation. Overall, SoZoGAN sets a new benchmark for turbulence super-resolution, providing an efficient, generalizable, and physically consistent platform for advancing turbulence modeling in computational fluid dynamics.

\backsection[Code availability]
{Sample codes for constituting and training our SoZoGAN framework are available at \href{https://github.com/HookGoh/Tur-SRGANs.git}{https://github.com/HookGoh/Tur-SRGANs.git}.}

\backsection[Acknowledgements]
{This study was supported by the National Natural Science Foundation of China (52478535), the National Key R\&D Program of China (2023YFE0120000), the Natural Science Foundation of Chongqing (CSTB2023NSCQ-MSX0060), the Future Foundation of the Energy Science (WLNY-MS-2022-014), and Guangdong Basic and Applied Basic Research Foundation (2024A1515240081).}

\backsection[Declaration of interests]
{The authors report no conflict of interest.}

\bibliographystyle{jfm}


\end{document}